\documentclass[12pt]{article}
\usepackage{amssymb, amsmath, bm, float, epsfig, color, slashed, dsfont, subcaption, multirow, arydshln}
\usepackage[centering]{geometry}
\makeatletter
\@addtoreset{equation}{section}
\makeatother

\begin{document}

\thispagestyle{empty}

\begin{flushright}
KOBE-TH-20-03 \\
DESY 20-063 \\
\end{flushright}
\vspace{40pt}
\begin{center}
{\Large\bf Zero-mode counting formula and zeros\\[5pt] in orbifold compactifications} \\

\vspace{40pt}
{\bf{Makoto Sakamoto$^\dagger\hspace{.5pt}$}\footnote{E-mail: dragon@kobe-u.ac.jp}}, \, 
{\bf{Maki Takeuchi$^\dagger\hspace{.5pt}$}\footnote{E-mail: 191s107s@stu.kobe-u.ac.jp}}, \, 
{\bf{Yoshiyuki Tatsuta$^\ast\hspace{.5pt}$}\footnote{E-mail: yoshiyuki.tatsuta@desy.de}} \\

\vspace{40pt}
{\it $^\dagger$ Department of Physics, Kobe University, Kobe 657-8501, Japan \\[5pt]
$^\ast$ Deutsches Elektronen-Synchrotron DESY, Hamburg 22607, Germany} \\
\end{center}

\vspace{30pt}
\begin{abstract}
\noindent
We thoroughly analyze the number of independent zero modes and their zero points
on the toroidal orbifold $T^2/\mathbb{Z}_N \,\, (N = 2, 3, 4, 6)$
with magnetic flux background, inspired by the Atiyah-Singer index theorem.
We first show a complete list for the number $n_{\eta}$ of orbifold zero modes belonging to $\mathbb{Z}_{N}$ eigenvalue $\eta$.
Since it turns out that $n_{\eta}$ quite complicatedly depends on the flux quanta $M$, the Scherk-Schwarz twist phase $(\alpha_1, \alpha_2)$, and the $\mathbb{Z}_{N}$ eigenvalue $\eta$, it seems hard that $n_{\eta}$
can be universally explained in a simple formula.
We, however, succeed in finding a single zero-mode counting  formula
$n_{\eta} = (M-V_{\eta})/N + 1$, where $V_{\eta}$ denotes the sum of
winding numbers at the fixed points on the orbifold $T^2/\mathbb{Z}_N$.
The formula is shown to hold for any pattern.
\end{abstract}

\newpage
\setcounter{page}{2}
\setcounter{footnote}{0}

\section{Introduction}
Over the long history of physics, the Atiyah-Singer index theorem \cite{Atiyah:1963zz} has played important roles.
After it was proposed in 1963, many applications have been done for physics.
The index theorem claims that the index of a Dirac operator $\slashed{D}$
\begin{gather}
{\rm Ind} \, (i\slashed{D}) \equiv n_{+} - n_{-}
\end{gather}
is a topological invariant.
Here, $n_{\pm}$ denotes the number of $\pm$ chiral zero modes for the Dirac
operator $\slashed{D}$.
Indeed, it is quite powerful to clearly extract some essential features.

There are various applications in physics.
One is the chiral anomaly in gauge theory.
The computation by use of the path integral \cite{Fujikawa:1979ay, Fujikawa:1980eg} 
can be mathematically justified by considering it as a special case of the theorem.
The second is the Witten index in supersymmetric theory \cite{Witten:1982df}.
The Witten index plays an important role in constructions of supersymmetric
models with spontaneous supersymmetry breaking,
because supersymmetry remains unbroken if the Witten index is non-vanishing.
The theorem has been applied to string theory in the context of flux compactifications \cite{Witten:1984dg, Bachas:1995ik}, where it has been used to count the number of chiral zero modes appearing in the four-dimensional (4d) effective (field) theory.

For both higher-dimensional field theory and string theory, a crucial difficulty to connect our world is to obtain chiral spectra.
A promising method to realize the chiral spectra has been known as magnetic flux compactifications in type-I and II string theory \cite{Abouelsaood:1986gd, Blumenhagen:2000wh, Angelantonj:2000hi, Angelantonj:2002ct, Blumenhagen:2005mu, Blumenhagen:2006ci, Ibanez:2012zz}.
The magnetic compactifications have provided semi-realistic models in the context of string phenomenology as well as at the field theory level,
e.g. three-generation models \cite{Abe:2008sx, Abe:2015yva}, flavor structures \cite{Abe:2014vza, Fujimoto:2016zjs, Kobayashi:2016qag, Buchmuller:2017vho, Buchmuller:2017vut}, and some applications to physics beyond the Standard Model \cite{Higaki:2016ydn, Buchmuller:2015jna, Buchmuller:2019ipg}.

On the two-dimensional (2d) torus $T^{2}$ with magnetic flux background,
the Atiyah-Singer index theorem is known as \cite{Witten:1984dg, Green:1987mn}
\begin{gather}
n_{+} - n_{-} = \frac{q}{2\pi} \int_{T^2} F = M,
\label{eq1.2}
\end{gather}
where $M$ is the flux quanta in the torus compactification.
Thus, the number of chiral zero modes is given by a simple formula (\ref{eq1.2}) on the torus. 
It is instructive to note that the index can be alternatively expressed by
counting winding numbers at zero points of zero mode wavefunctions
\cite{VENUGOPALKRISHNA1972349, aliprantis2002invitation}.

The number of chiral zero modes on the magnetized orbifold
$T^{2}/\mathbb{Z}_{N}$ ($N=2,3,4,6$) has been explored in 
\cite{Abe:2013bca, Abe:2014noa, Kobayashi:2017dyu}.
However, a list of chiral zero-mode numbers on the orbifolds has not been completed, due to its complicated dependence on the flux quanta $M$, the Scherk-Schwarz (SS) twist phase $(\alpha_{1}, \alpha_{2})$, 
and the $\mathbb{Z}_{N}$ eigenvalue $\eta$ under the $\mathbb{Z}_{N}$ rotation.
Furthermore, unlike the index theorem on the torus, any simple formula has not been known for the number of zero modes on the orbifolds.

One of our goals in this paper is to give a complete list of $\mathbb{Z}_{N}$ zero-mode numbers on the orbifold $T^{2}/\mathbb{Z}_{N}$ ($N=2,3,4,6$).
This is the main subject in Section 3, and the list is given in Tables 1\,--\,4.
The other is to find a zero-mode counting formula by which all the $\mathbb{Z}_{N}$ zero-mode numbers can be counted universally.

We actually claim the following zero-mode counting formula\footnote{
In this paper, we will mainly concentrate on the case of $M > 0$,
for which there is no negative chiral zero mode, i.e. $n_{-}=0$.}
\begin{gather}
n_{\eta} = \frac{M-V_{\eta}}{N} + 1,
\label{eq1.3}
\end{gather}
where $n_{\eta}$ denotes the number of zero modes belonging to the
$\mathbb{Z}_{N}$ eigenvalue $\eta$, and $V_{\eta}$ is the sum of 
winding numbers associated with zeros at the fixed points on the
orbifold $T^{2}/\mathbb{Z}_{N}$.
In Section 4, we verify that the formula (\ref{eq1.3}) really holds for any of
the flux quanta, the SS twist phase, and the $\mathbb{Z}_{N}$ eigenvalue.
It is the most important result in this paper.

This paper is organized as follows. 
In Section 2, we briefly review zero modes on the orbifold $T^2/\mathbb{Z}_N \,\, (N=2, 3, 4, 6)$.
In Section 3, we claim the number of independent orbifold zero modes for arbitrary $M$.
In Section 4, inspired by the Atiyah-Singer index theorem, we explore a formula that uniquely tells the number of orbifold zero modes.
Section 5 is devoted to discussion and conclusion.
In appendices, we mention our notation and also derive a formula used in our discussions.

\section{Zero modes on orbifolds}
In this section, we briefly review zero mode wavefunctions on 2d toroidal orbifold
$T^{2}/\mathbb{Z}_{N}$ ($N=2,3,4,6$) with magnetic flux background 
\cite{Abe:2013bca, Abe:2014noa}.

\subsection{Abelian six-dimensional gauge theory}
In this paper, we consider a six-dimensional (6d) gauge theory compactified
on $T^{2}$ or $T^{2}/\mathbb{Z}_{N}$.
Using the complex coordinate $z \equiv y_{1} + \tau y_{2}$, the torus $T^{2}$
is obtained by the identification 
$z \sim z+1 \sim z+\tau$ ($\tau \in \mathbb{C}, \, \textrm{Im}\,\tau > 0$)
under torus lattice shifts.

Following \cite{Cremades:2004wa}, we assume a non-trivial magnetic flux background in the (1-form) vector potential:
\begin{gather}
A(z) \equiv \frac{f}{2 \, {\rm Im} \, \tau} {\rm Im} \, (\bar z dz ),
\label{eq2.1}
\end{gather}
where $f$ denotes the homogeneous flux on the torus.
Torus lattice shifts on the vector potential should be accompanied by
gauge transformation
\begin{gather}
A (z + 1) = A (z) + d \Lambda_1(z), \label{BCforA1}\\
A (z + \tau) = A (z) + d \Lambda_2(z),  \label{BCforA2}
\end{gather}
where $\Lambda_1(z)$ and $\Lambda_2(z)$ are gauge parameters given by
\begin{gather}
\Lambda_1(z) = \frac{f}{2 \, {\rm Im} \, \tau} {\rm Im} \, z, \qquad \Lambda_2(z) = \frac{f}{2 \, {\rm Im} \, \tau} {\rm Im} \, (\bar \tau z).
\end{gather}
It is shown in \cite{Abe:2013bca} that Wilson lines can be set to be vanishing 
without loss of generality, and we do not treat them in the following.
The background vector potential (\ref{eq2.1}) leads to a non-trivial background 
of the (2-form) field strength $F$ such that $\int_{T^{2}} F = f$.

Next, we look at a 6d Weyl fermion in the flux background.
The Lagrangian reads
\begin{gather}
{\cal L}_{\rm 6d} = i \bar \Psi \Gamma^M D_M \Psi, \qquad \Gamma_7 \Psi = \Psi,
\end{gather}
where $M \,\, (= 0,1,2,3,5,6)$ is the 6d spacetime index, and 
$\Gamma^0, \Gamma^1, \ldots, \Gamma^6$ denote 6d gamma matrices.
$\Gamma_7$ denotes the 6d chirality operator and $D_M = \partial_M - iq A_M$ is 
the covariant derivative.
The 6d Weyl fermion $\Psi(x, z)$ can be decomposed into
4d Weyl left/right-handed fermions $\psi^{(4)}_{\textrm{L/R}}(x)$ as
\begin{gather}
\Psi(x, z) = \sum_{n, \hspace{.5pt} j} 
\bigl( \psi^{(4)}_{\textrm{R}, \hspace{.5pt} n, \hspace{.5pt} j}(x) \otimes \psi^{(2)}_{+, \hspace{.5pt} n, \hspace{.5pt} j}(z) + \psi^{(4)}_{\textrm{L}, \hspace{.5pt} n, \hspace{.5pt} j}(x) \otimes \psi^{(2)}_{-, \hspace{.5pt} n, \hspace{.5pt} j}(z) \bigr),
\end{gather}
where $x^{\mu} \,\, (\mu=0,1,2,3)$ denotes the 4d Minkowski coordinate.
For convenience, we adopt the following notation for 2d Weyl fermions:
\begin{gather}
\psi^{(2)}_{+, \hspace{.5pt} n, \hspace{.5pt} j} = 
\begin{pmatrix}
\psi_{+, \hspace{.5pt} n, \hspace{.5pt} j} \\[3pt] 0
\end{pmatrix}
, \qquad \psi^{(2)}_{-, \hspace{.5pt} n, \hspace{.5pt} j} = 
\begin{pmatrix}
0 \\[3pt] \psi_{-, \hspace{.5pt} n, \hspace{.5pt} j}
\end{pmatrix},
\label{psipm}
\end{gather}
where $n$ and $j$ label each of the Landau level and the degeneracy of mode functions 
on each level, respectively.

The 2d Weyl fermions are required to satisfy the pseudo-periodic boundary conditions associated with the gauge transformation:
\begin{gather}
\psi_{\pm, \hspace{.5pt} n, \hspace{.5pt} j}(z + 1) = U_1(z) \psi_{\pm, \hspace{.5pt} n, \hspace{.5pt} j}(z), \qquad \psi_{\pm, \hspace{.5pt} n, \hspace{.5pt} j}(z + \tau) = U_2(z) \psi_{\pm, \hspace{.5pt} n, \hspace{.5pt} j}(z)
\label{BCs}
\end{gather}
with
\begin{gather}
U_i(z) = e^{i q \Lambda_i (z)} e^{2\pi i \alpha_i} \quad (i=1,2),
\end{gather}
and $\alpha_i \,\, (i=1,2)$ corresponds to the Scherk-Schwarz twist phase.

As claimed in \cite{Bachas:1995ik, Abouelsaood:1986gd}, 
the gauge transformation above is well-defined on the torus if and only if the homogeneous flux $f$ is quantized as
\begin{gather}
\frac{qf}{2 \pi} \equiv M \in \mathbb{Z}.
\label{fluxquantization}
\end{gather}
When going to toroidal orbifolds, one has to be careful of the localized fluxes at orbifold fixed points.
By computing Wilson loops around the fixed points, one finds that, in general, there exist the non-zero contributions of the localized fluxes on the orbifolds \cite{Buchmuller:2015eya}.
Then, taking into account all the localized fluxes, it can be confirmed that the flux quantization condition \eqref{fluxquantization} is available on $T^2/\mathbb{Z}_N$ as well \cite{Buchmuller:2018lkz}.

\subsection{Zero modes on $T^2$}
In this subsection, we show zero mode wavefunctions on the torus $T^{2}$ \cite{Cremades:2004wa}.
To make our analysis simple, we restrict ourselves to $M>0$,
although one can analyze the case of $M<0$ in a similar way.

Focusing on the lowest-lying states $n = 0$, we omit such an index in what follows.
Zero mode equations are found as
\begin{gather}
\left(\bar \partial + \frac{\pi M}{2 \, {\rm Im}\, \tau} z \right) \psi_{+, \hspace{.5pt} j}(z) = 0, \qquad \left(\partial - \frac{\pi M}{2 \, {\rm Im}\, \tau} \bar z \right) \psi_{-, \hspace{.5pt} j}(z) = 0.
\end{gather}
Imposing the boundary conditions \eqref{BCs}, we find $M$-fold normalizable
zero mode solutions only for $\psi_{+}$,\footnote{For $M<0$, there exist $|M|$-fold normalizable zero mode solutions only for $\psi_{-}$.} i.e.
\begin{align}
\psi_{+, \hspace{.5pt} j}(z) &= \mathcal{N} e^{i \pi M z \, {\rm Im}\,z / {\rm Im}\, \tau} \, \vartheta
\begin{bmatrix}
\tfrac{j + \alpha_1}M \\[3pt] -\alpha_2
\end{bmatrix}
(Mz, M\tau) \notag\\
&\equiv \xi^j(z).
\label{toruszeromodes}
\end{align}
Here, $j = 0, 1, \ldots, M-1$ stand for the degeneracy of zero mode solutions, 
and ${\cal N}$ is a normalization constant determined by
\begin{gather}
\int_{T^2} d^2z \, \xi^j(z) \bigl(\xi^k(z) \bigr)^* = \delta_{j, \hspace{.5pt} k}.
\end{gather}
The Jacobi $\vartheta$-function is defined by
\begin{align}
\vartheta
\begin{bmatrix}
a \\[3pt] b
\end{bmatrix}
(c, d)
 = \sum_{l=-\infty}^{\infty}
   e^{\pi i(a+l)^{2}d}\,e^{2\pi i(a+l)(c + b)}.
\label{thetafunction}
\end{align}

The result \eqref{toruszeromodes} immediately implies that the flux quanta $M$ lead to $M$-fold 4d chiral Weyl fermions 
$\psi^{(4)}_{\textrm{R}, \hspace{.5pt} 0, \hspace{.5pt} j}(x) \,\, (j=0,1, \ldots ,M-1)$.
Notice that the zero mode wavefunctions $\xi^{j}(z)$ are characterized by the flux quanta $M$ 
and the SS twist phase $(\alpha_1, \alpha_2)$.
For later convenience, it is useful to schematically express 
the zero mode wavefunctions as
\begin{gather}
\xi^j (z) \equiv \langle z \hspace{.5pt} | \hspace{.5pt} M, j, \alpha_1, \alpha_2 \rangle_{T^2}.
\label{torusphsyicalstate}
\end{gather}
Hereafter, we call $| \hspace{.5pt} M, j, \alpha_1, \alpha_2 \rangle_{T^2}$ {\em torus physical states}.

\subsection{Zero modes on $T^2/\mathbb{Z}_N$}
We now move on to the orbifold $T^{2}/\mathbb{Z}_{N} \,\, (N=2,3,4,6)$,
which is our main subject in this paper.
The orbifold $T^2/\mathbb{Z}_N$ is given by the torus identification 
and an additional $\mathbb{Z}_N$ one
\begin{gather}
z \sim \omega z \qquad (\omega \equiv e^{2 \pi i/N}).
\label{orbifolding}
\end{gather}

As discussed in \cite{Choi:2006qh} from the viewpoint of crystallography, 
we first need to clarify a relation between $\omega$ and a complex modulus $\tau$.
For $N=2$, $\tau$ is arbitrary as long as ${\rm Im} \, \tau > 0$.
For $N = 3, 4, 6$, we must impose $\tau = \omega \,\, (= e^{2 \pi i/N})$.
The orbifold fixed points, which are invariant under the $\mathbb{Z}_{N}$ rotations up to torus lattice shifts, are found as
\begin{align}
(y_{1},y_{2})
 = 
\begin{cases}
(0,0), (1/2, 0), (0, 1/2), (1/2, 1/2) & \quad \textrm{on} ~ T^{2}/\mathbb{Z}_{2},\\ 
(0,0), (2/3, 1/3), (1/3, 2/3) & \quad \textrm{on} ~ T^{2}/\mathbb{Z}_{3},\\ 
(0,0), (1/2, 1/2) & \quad \textrm{on} ~ T^{2}/\mathbb{Z}_{4},\\
(0,0) & \quad \textrm{on} ~ T^{2}/\mathbb{Z}_{6}.
\end{cases}
\label{fixedpoint}  
\end{align}

To be consistent with the orbifold identification, the SS twist phase
($\alpha_{1}, \alpha_{2}$) turns out to be quantized as
\begin{gather}
(\alpha_{1}, \alpha_{2})
 = (0,0), (1/2,0), (0,1/2), (1/2,1/2) \qquad \textrm{on} ~ T^{2}/\mathbb{Z}_{2},
\label{Z2_SSphase}  \\
\alpha = \alpha_{1} = \alpha_{2} = 
\begin{cases} 
0, 1/3, 2/3 & \quad (M = \textrm{even})\\
1/6, 3/6, 5/6 & \quad (M = \textrm{odd})
\end{cases} \qquad \textrm{on} ~ T^2/\mathbb{Z}_3,
\label{Z3_SSphase} \\
\alpha = \alpha_{1} = \alpha_{2}
 = 0, 1/2 \qquad \textrm{on} ~ T^{2}/\mathbb{Z}_{4},
\label{Z4_SSphase} \\
\alpha = \alpha_{1} = \alpha_{2} = 
\begin{cases} 
0 & \quad (M = \textrm{even})\\
1/2 & \quad (M = \textrm{odd})
\end{cases} \qquad \mathrm{on} ~ T^2/\mathbb{Z}_6.
\label{Z6_SSphase}
\end{gather}

Wavefunctions on the orbifold $T^{2}/\mathbb{Z}_{N}$ are classified by 
$\mathbb{Z}_{N}$ eigenvalues under the $\mathbb{Z}_{N}$ rotation
$z \to \omega z$ as
\begin{gather}
\psi_{+, \hspace{.5pt} n, \hspace{.5pt} j}(\omega z) = \eta \, \psi_{+, \hspace{.5pt} n, \hspace{.5pt} j}(z), \qquad \psi_{-, \hspace{.5pt} n, \hspace{.5pt} j}(\omega z) = \omega \eta \, \psi_{-, \hspace{.5pt} n, \hspace{.5pt} j}(z),
\label{ZNparity}
\end{gather}
where $\eta = \omega^{\ell} \,\, (\ell=0,1,\ldots, N-1)$ denotes the $\mathbb{Z}_{N}$ eigenvalue.

Again, we focus only on $M>0$ and the $\mathbb{Z}_N$ eigenstates for $\psi_{+}$ satisfying \eqref{BCs} and \eqref{ZNparity}. 
In terms of the zero mode wavefunctions $\xi^{j}(z)$ on $T^{2}$,
formal solutions to \eqref{ZNparity} are constructed as
\begin{gather}
\xi^j_{\eta} (z) = \mathcal{N}^{j}_{\eta} \, \sum_{\ell = 0}^{N-1} \bar \eta^\ell \, \xi^j(\omega^\ell z)
\qquad (\eta = 1,\omega,\ldots,\omega^{N-1}),
\end{gather}
where $\mathcal{N}^{j}_{\eta}$ is a normalization constant and not relevant for our discussions.
A difficulty is that all the eigen wavefunctions $\xi^{j}_{\eta}(z) \,\, (j=0,1,\ldots,M-1)$ are not always linearly independent.

One of our goals in this paper is to find the number of independent 
$\mathbb{Z}_{N}$ eigenstates for each $\mathbb{Z}_{N}$ eigenvalue $\eta$.
In \cite{Abe:2013bca}, for some small values of $M$, the number of 
independent $\mathbb{Z}_{N}$ eigenstates has been obtained.
It is, however, difficult to find the number of them for large $M$
(except on $T^2/\mathbb{Z}_{2}$).

Another way to obtain the number of independent $\mathbb{Z}_{N}$ eigenstates is to use a property of the torus physical states 
$| \hspace{.5pt} M, k, \alpha_1, \alpha_2 \rangle_{T^2}$ under the $\mathbb{Z}_{N}$ rotation:
\begin{gather}
\hat U_{\mathbb{Z}_{\hspace{-.5pt}N}} | \hspace{.5pt} M, j, \alpha_1, \alpha_2 \rangle_{T^2} 
= \sum_{k = 0}^{M-1} D_{jk} \hspace{.5pt} | \hspace{.5pt} M, k, \alpha_1, \alpha_2 \rangle_{T^2} \quad (j=0,1, ..., M-1),
\label{matD}
\end{gather}
where $\hat U_{\mathbb{Z}_{\hspace{-.5pt}N}}$ is the $\mathbb{Z}_{N}$ rotation operator.
We summarize the results of $D_{jk}$ \cite{Abe:2014noa}:\footnote{
We understand our definition of the Kronecker delta as
\begin{gather}
\delta_{j, \hspace{.5pt} k} =
\begin{cases}
1 & \quad (j = k ~~ {\rm mod}~M), \\
0 & \quad (j \neq k ~~ {\rm mod}~M).
\end{cases}
\end{gather}
}
\begin{gather}
D_{jk} = 
\begin{cases}
\displaystyle e^{-2 \pi i (j + \alpha_1) \frac{2 \alpha_2}{M}} \, \delta_{-2\alpha_1-j, \hspace{.5pt} k}
& \quad \textrm{for} ~ T^{2}/\mathbb{Z}_{2}, \\[3pt]
\displaystyle \frac1{\sqrt M} \, e^{-i \frac{\pi}{12} + i \frac{3 \pi \alpha^2}{M}} e^{i \frac{\pi}{M} k(k + 6\alpha) + 2\pi i \frac{jk}{M}} 
& \quad \textrm{for} ~ T^{2}/\mathbb{Z}_{3}, \\[10pt]
\displaystyle \frac1{\sqrt M} \, e^{2 \pi i \frac{\alpha^2}M} e^{2 \pi i \frac{jk}M + 2 \pi i \frac{2\alpha}M k}
& \quad \textrm{for} ~ T^{2}/\mathbb{Z}_{4}, \\[10pt]
\displaystyle \frac1{\sqrt M} \, e^{i \frac{\pi}{12} + i \frac{\pi \alpha^2}{M}} e^{- i \frac{\pi}{M} k^2 + 2 \pi i \frac{\alpha}M k + 2\pi i \frac{jk}{M}}
& \quad \textrm{for} ~ T^{2}/\mathbb{Z}_{6}.
\label{D_{jk}}
\end{cases}
\end{gather}

The number of independent $\mathbb{Z}_{N}$ eigenstates can be obtained
by analyzing eigenvalues of the $M$-by-$M$ matrix $D_{jk}$.
Since $(\hat{U}_{\mathbb{Z}_{N}})^{N} = \mathds{1}$, the eigenvalues of 
$D_{jk}$ are $1,\omega,\ldots,\omega^{N-1} \,\, (\omega = e^{2\pi i/N})$,
and the degeneracy of each eigenvalue corresponds to the number of independent
$\mathbb{Z}_{N}$ eigenstates.
Thus, it could be, in principle, obtained by diagonalizing the
$M$-by-$M$ matrix $D_{jk}$.
In \cite{Abe:2014noa}, for some small $M$, the number of 
independent $\mathbb{Z}_{N}$ eigenstates has been obtained and
found to agree with the previous results given in \cite{Abe:2013bca}.
The authors have not been, however, succeeded in deriving a general list for the numbers of independent $\mathbb{Z}_{N}$ eigenstates.

In the next section, we analyze each eigenvalue of the matrix
$D_{jk}$ and give a complete list for the numbers of the $\mathbb{Z}_{N}$ eigenstates for any of the flux quanta $M$, 
the SS twist phase $(\alpha_{1},\alpha_{2})$, and the $\mathbb{Z}_{N}$ 
eigenvalue $\eta$.

\section{Counting independent $\mathbb{Z}_{N}$ eigenstates}
The numbers of $\mathbb{Z}_{N}$ eigen zero modes have been obtained on $T^{2}/\mathbb{Z}_{2}$ for arbitrary $M$ and on $T^{2}/\mathbb{Z}_{N}$ ($N=3,4,6$) for
some small $M$ in \cite{Abe:2013bca, Abe:2014noa}.
There is another way to discuss orbifold zero modes by use of modular
transformations \cite{Kobayashi:2017dyu}.
Nevertheless, unclear is how to introduce non-zero SS twist phases. 
In this section, we give a complete list for the numbers of 
$\mathbb{Z}_{N}$ eigen zero modes on all the orbifolds
$T^{2}/\mathbb{Z}_{N} \,\, (N=2,3,4,6)$ for any of the flux quanta $M \, (>0)$,
the SS twist phases, and the $\mathbb{Z}_{N}$ eigenvalues.
It is one of our main results in this paper.

\subsection{$T^2/\mathbb{Z}_2$}
We start by considering the $\mathbb{Z}_2$ transformation property,
\begin{gather}
\hat U_{\mathbb{Z}_2} | \hspace{.5pt} M, j, \alpha_1, \alpha_2 \rangle_{T^2} = \sum_{k = 0}^{M-1} D_{jk}(\alpha_1, \alpha_2) \hspace{.5pt} | \hspace{.5pt} M, k, \alpha_1, \alpha_2 \rangle_{T^2},
\end{gather}
where $D_{jk}(\alpha_{1},\alpha_{2})$ is given in \eqref{D_{jk}}.
For later convenience, we have explicitly written down the SS-phase dependence $(\alpha_1, \alpha_2)$ within $D_{jk}$.
Due to $(\hat U_{\mathbb{Z}_2})^2 = \mathds{1}$, the $M$-by-$M$ matrix $D_{jk}(\alpha_1, \alpha_2)$ gives eigenvalues $\pm 1$.
Then, the number of $\pm 1$ eigenvalues corresponds to the number of orbifold physical states belonging to $\mathbb{Z}_2$ eigenvalues $\eta = \pm 1$.
We now find
\begin{gather}
{\rm tr} \, \bigl(D (\alpha_1, \alpha_2) \bigr) = n_+ - n_-,
\end{gather}
where we define $n_\pm$ as the number of orbifold physical states 
with $\mathbb{Z}_2$ eigenvalues $\eta = \pm 1$.
Moreover, $n_\pm$ must satisfy
\begin{gather}
n_+ + n_- = M.
\label{Z2_relation}
\end{gather}

\subsubsection*{\underline{$(\alpha_1, \alpha_2) = (0,0)$}}
Using \eqref{Z2_relation} and the relation
\begin{gather}
n_+ - n_- = {\rm tr} \, \bigl(D (0,0) \bigr) = 
\begin{cases}
1 & \quad (M = 2m+1), \\
2 & \quad (M = 2m+2),
\label{eq3.4}
\end{cases}
\end{gather}
with $m \in \mathbb{N} \cup \{ 0 \}$, we easily obtain
\begin{gather}
n_+ = \frac{M+1}2, \qquad n_- = \frac{M-1}2 \qquad (M = 2m+1), \\
n_+ = \frac{M}2 + 1, \qquad n_- = \frac{M}2 - 1 \qquad (M = 2m+2).
\end{gather}
Here, we have used the expression \eqref{D_{jk}} in the last equality of 
\eqref{eq3.4}.
These results are summarized in Table \ref{tab_Z2} (a).

\subsubsection*{\underline{$(\alpha_1, \alpha_2) = (1/2,0)$}}
Similarly, using 
\begin{gather}
n_+ - n_- = {\rm tr}\,\bigl(D (\tfrac12, 0) \bigr) = 
\begin{cases}
1 & \quad (M=2m+1), \\
0 & \quad (M=2m+2),
\end{cases}
\end{gather}
we obtain
\begin{gather}
n_+ = \frac{M+1}2, \qquad n_- = \frac{M-1}2 \qquad (M=2m+1), \\
n_+ = n_- = \frac{M}2 \qquad (M=2m+2).
\end{gather}
These results are summarized in Table \ref{tab_Z2} (b).

\subsubsection*{\underline{$(\alpha_1, \alpha_2) = (0,1/2)$}}
Similarly, using 
\begin{gather}
n_+ - n_- = {\rm tr}\,\bigl(D (0, \tfrac12) \bigr) = 
\begin{cases}
1 & \quad (M=2m+1), \\
0 & \quad (M=2m+2),
\end{cases}
\end{gather}
we obtain
\begin{gather}
n_+ = \frac{M+1}2, \qquad n_- = \frac{M-1}2 \qquad (M=2m+1), \\
n_+ = n_- = \frac{M}2 \qquad (M=2m+2).
\end{gather}
These results are summarized in Table \ref{tab_Z2} (c).

\subsubsection*{\underline{$(\alpha_1, \alpha_2) = (1/2,1/2)$}}
Similarly, once more using 
\begin{gather}
n_+ - n_- = {\rm tr}\,\bigl(D (\tfrac12, \tfrac12) \bigr) = 
\begin{cases}
-1 & \quad (M=2m+1), \\
0 & \quad (M=2m+2),
\end{cases}
\end{gather}
we obtain
\begin{gather}
n_+ = \frac{M-1}2, \qquad n_- = \frac{M+1}2 \qquad (M=2m+1), \\
n_+ = n_- = \frac{M}2 \qquad (M=2m+2).
\end{gather}
These results are summarized in Table \ref{tab_Z2} (d).

\begin{table}[t]
\centering
{\tabcolsep = 4mm
\renewcommand{\arraystretch}{1.5}
\begin{tabular}{cc}
\begin{tabular}{c|cc}\hline
 & $M = 2m+1$ & $M = 2m+2$ \\ \hline
$n_+$ & $\tfrac{M+1}2$ & $\tfrac{M}2 + 1$ \\
$n_-$ & $\tfrac{M-1}2$ & $\tfrac{M}2 - 1$ \\ \hline
\end{tabular} &
\begin{tabular}{c|cc}\hline
 & $M = 2m+1$ & $M = 2m+2$ \\ \hline
$n_+$ & $\tfrac{M+1}2$ & $\tfrac{M}2$ \\
$n_-$ & $\tfrac{M-1}2$ & $\tfrac{M}2$ \\ \hline
\end{tabular} \\
(a) ~ $(\alpha_1, \alpha_2)=(0,0)$ & (b) ~ $(\alpha_1, \alpha_2)=(1/2,0)$ \\
 & \\
\begin{tabular}{c|cc}\hline
 & $M = 2m+1$ & $M = 2m$+2 \\ \hline
$n_+$ & $\tfrac{M+1}2$ & $\tfrac{M}2$ \\
$n_-$ & $\tfrac{M-1}2$ & $\tfrac{M}2$ \\ \hline
\end{tabular} & 
\begin{tabular}{c|cc}\hline
 & $M = 2m+1$ & $M = 2m+2$ \\ \hline
$n_+$ & $\tfrac{M-1}2$ & $\tfrac{M}2$ \\
$n_-$ & $\tfrac{M+1}2$ & $\tfrac{M}2$ \\ \hline
\end{tabular} \\
(c) ~ $(\alpha_1, \alpha_2)=(0,1/2)$ & (d) ~ $(\alpha_1, \alpha_2)=(1/2,1/2)$ \\
\end{tabular}
}
\caption{The number of independent physical zero modes on $T^2/\mathbb{Z}_2$.}
\label{tab_Z2}
\end{table}

\subsection{$T^2/\mathbb{Z}_3$}
We now move to $T^2/\mathbb{Z}_3$ and start with the $\mathbb{Z}_3$ transformation property,
\begin{gather}
\hat U_{\mathbb{Z}_3} | \hspace{.5pt} M, j, \alpha, \alpha \rangle_{T^2} = \sum_{k = 0}^{M-1} D_{jk}(\alpha) \hspace{.5pt} | \hspace{.5pt} M, k, \alpha, \alpha \rangle_{T^2},
\end{gather}
where $D_{jk}(\alpha) \equiv D_{jk}(\alpha,\alpha)$ is given in \eqref{D_{jk}}.
For later convenience, we have explicitly written down the SS-phase dependence $\alpha$ within $D_{jk}$.
Due to $(\hat U_{\mathbb{Z}_3})^3 = \mathds{1}$, the $M$-by-$M$ matrix $D_{jk}$ gives eigenvalues $1, \omega, \omega^2 \,\, (\omega = e^{2\pi i/3})$.
In analogy to $T^2/\mathbb{Z}_2$, we now find
\begin{align}
{\rm tr} \, \bigl(D (\alpha) \bigr) &= n_1 + \omega n_\omega + \omega^2 n_{\omega^2} \notag \\
&= n_1 - n_{\omega^2} + \omega(n_\omega - n_{\omega^2}),
\end{align}
where we have used $1 + \omega + \omega^2 = 0$.
We again define $n_{1, \hspace{1pt} \omega, \hspace{1pt} \omega^2}$ as the number of $\mathbb{Z}_{3}$ eigenstates belonging to $\mathbb{Z}_3$ eigenvalue $\eta = 1, \omega, \omega^2$, respectively.
Moreover, $n_{1, \hspace{1pt} \omega, \hspace{1pt} \omega^2}$ must satisfy
\begin{gather}
n_1 + n_\omega + n_{\omega^2} = M.
\label{Z3_relation}
\end{gather}

To derive $n_{1, \hspace{1pt} \omega, \hspace{1pt} \omega^2}$ analytically, 
we need to evaluate the trace of $D(\alpha)$, i.e.
\begin{gather}
{\rm tr}\, \bigl( D(\alpha) \bigr) = \frac{e^{-i \pi/12
}}{\sqrt M} \, \sum_{k = 0}^{M-1} e^{i \frac{3 \pi}M (k + \alpha)^2}.
\label{Z3_Modd}
\end{gather}
To perform the sum over $k$ in the case of the trivial SS twist phase ($\alpha = 0$),
we will use the formula
\begin{align}
\frac1{\sqrt p} \sum_{n = 0}^{p-1} \exp \left( \frac{2 \pi i n^2 q}{p} \right) &= \frac{e^{i\pi/4}}{\sqrt{2q}} \sum_{n = 0}^{2q-1} \exp\left( -\frac{\pi i n^2 p}{2q} \right)
\label{LS1}
\end{align}
or its complex conjugation
\begin{align}
\frac1{\sqrt p} \sum_{n = 0}^{p-1} \exp \left( - \frac{2 \pi i n^2 q}{p} \right) &= \frac{e^{-i\pi/4}}{\sqrt{2q}} \sum_{n = 0}^{2q-1} \exp\left( \frac{\pi i n^2 p}{2q} \right)
\label{LS2}
\end{align}
for $p, q \in \mathbb{N}$.
These formulae are mathematically known as the Landsberg-Schaar relation.
Furthermore, for non-trivial SS twist phases ($\alpha \ne 0$), we need to use an extension of the formula
\begin{gather}
\frac1{\sqrt p} \sum_{n = 0}^{p-1} \exp \left( \frac{\pi i (n + \nu)^2 q}{p} \right) = \frac{e^{i\pi/4}}{\sqrt{q}} \sum_{n = 0}^{q-1} \exp\left( -\frac{\pi i n^2 p}{q} - 2 \pi i n \nu \right) \label{LS3}
\end{gather}
for $p, q \in \mathbb{N}$, $\nu \in \mathbb{Q}$, and $pq + 2q \nu \in 2\mathbb{Z}$.
Since that seems to be unfamiliar in physics,
we give an elementary proof of the generalized Landsberg-Schaar relation \eqref{LS3} in Appendix B.
As we will see below, it is interesting that the necessary condition
$pq + 2q\nu \in 2\mathbb{Z}$ is consistent with the allowed SS twist phases
\eqref{Z2_SSphase}\,--\,\eqref{Z6_SSphase}.

\subsubsection*{\underline{$\alpha = 0$}}
In this case, $M$ must be an even (positive) integer, as mentioned in Section 2.
Utilizing \eqref{LS2} with $p = 3$ and $2q = M$, we find
\begin{gather}
n_1 - n_{\omega^2} + \omega(n_\omega - n_{\omega^2}) = {\rm tr}\, \bigl(D(0) \bigr) =
\begin{cases}
-\omega & \quad (M = 6m + 2), \\
\omega & \quad (M = 6m + 4), \\
2 + \omega & \quad (M = 6m + 6),
\end{cases}
\label{trD_Z3_1}
\end{gather}
where $m \in \mathbb{N} \cup \{0\}$.
From \eqref{Z3_relation} and \eqref{trD_Z3_1}, we explicitly obtain
\begin{gather}
n_1 =  n_{\omega^2} = \frac{M+1}3, \qquad n_\omega = \frac{M-2}3, \qquad (M=6m+2), \\
n_1 = n_{\omega^2} = \frac{M-1}3, \qquad n_\omega = \frac{M+2}3, \qquad (M=6m+4), \\
n_1 = \frac{M}3 + 1, \qquad n_\omega = \frac{M}3, \qquad n_{\omega^2} = \frac{M}3 -1 \qquad (M=6m+6),
\end{gather}
as summarized in Table \ref{tab_Z3} (a).

\subsubsection*{\underline{$\alpha = 1/3, 2/3$}}
In this case, $M$ must be again an even (positive) integer. 
To evaluate \eqref{Z3_Modd}, we use the formula \eqref{LS3} 
for $p = M, q = 3$, and $\nu=1/3, 2/3$ (with $pq+2q\nu \in 2\mathbb{Z}$ satisfied).
Then, we find
\begin{align}
n_1 - n_{\omega^2} + \omega(n_\omega - n_{\omega^2}) &= {\rm tr}\, \bigl(D(1/3) \bigr) \notag \\
&= {\rm tr}\, \bigl(D(2/3) \bigr) \notag \\
&=
\begin{cases}
1+\omega & \quad (M = 6m + 2), \\
1 & \quad (M = 6m + 4), \\
0 & \quad (M = 6m + 6),
\end{cases}
\label{trD_Z3_2}
\end{align}
where $m \in \mathbb{N} \cup \{0\}$.
Similarly, by use of \eqref{Z3_relation} and \eqref{trD_Z3_2}, we explicitly obtain
\begin{gather}
n_1 =  n_\omega = \frac{M+1}3, \qquad n_{\omega^2} = \frac{M-2}3 \qquad (M=6m+2), \\
n_1 = \frac{M+2}3, \qquad n_\omega = n_{\omega^2} = \frac{M-1}3 \qquad (M=6m+4), \\
n_1 = n_\omega = n_{\omega^2} = \frac{M}3 \qquad (M=6m+6),
\end{gather}
as summarized in Table \ref{tab_Z3} (b).

\subsubsection*{\underline{$\alpha = 1/6, 5/6$}}
In this case, $M$ must be an odd (positive) integer, as mentioned in Section 2.
To evaluate \eqref{Z3_Modd}, we need to use the generalized relation \eqref{LS3} for $p = M$, $q = 3$, and $\nu = \alpha$ (with $pq + 2q \nu \in 2\mathbb{Z}$ satisfied).
Thus, it is straightforward to find
\begin{align}
n_1 - n_{\omega^2} + \omega(n_\omega - n_{\omega^2}) &= {\rm tr}\, \bigl(D(1/6) \bigr) \notag \\
&= {\rm tr}\, \bigl(D(5/6) \bigr) \notag \\
&=
\begin{cases}
1 & \quad (M = 6m+1), \\
0 & \quad (M = 6m+3), \\
1 + \omega & \quad (M = 6m+5),
\end{cases}
\end{align}
where $m \in \mathbb{N} \cup \{0\}$.
These equations immediately lead to 
\begin{gather}
n_1 = \frac{M+2}3, \qquad n_\omega = n_{\omega^2} = \frac{M-1}3 \qquad (M=6m+1), \\[5pt]
n_1 =  n_\omega = n_{\omega^2} = \frac{M}3 \qquad (M=6m+3), \\[5pt]
n_1 = n_\omega = \frac{M+1}3, \qquad n_{\omega^2} = \frac{M-2}3 \qquad (M=6m+5),
\end{gather}
as summarized in Table \ref{tab_Z3} (c).

\subsubsection*{\underline{$\alpha = 1/2$}}
Similarly, using 
\begin{gather}
n_1 - n_{\omega^2} + \omega(n_\omega - n_{\omega^2}) = {\rm tr}\, \bigl(D(1/2) \bigr) = 
\begin{cases}
\omega & \quad (M = 6m+1), \\
2 + \omega & \quad (M = 6m+3), \\
-\omega & \quad (M = 6m+5),
\end{cases}
\end{gather}
where $m \in \mathbb{N} \cup \{0\}$,
we easily reach
\begin{gather}
n_1 =  n_{\omega^2} = \frac{M-1}3, \qquad n_\omega = \frac{M+2}3, \qquad (M=6m+1), \\[5pt]
n_1 =  \frac{M}3 + 1, \qquad n_\omega = \frac{M}3, \qquad n_{\omega^2} = \frac{M}3 -1 \qquad (M=6m+3), \\[5pt]
n_1 = n_{\omega^2} = \frac{M+1}3, \qquad n_\omega = \frac{M-2}3 \qquad (M=6m+5),
\end{gather}
as summarized in Table \ref{tab_Z3} (d).

\begin{table}[!tb]
\centering
{\tabcolsep = 4mm
\renewcommand{\arraystretch}{1.5}
\begin{subtable}{\textwidth}\centering
\begin{tabular}{c|ccc}\hline
 & $M = 6m+2$ & $M = 6m+4$ & $M = 6m + 6$ \\ \hline
$n_1$ & $\tfrac{M+1}3$ & $\tfrac{M-1}3$ & $\tfrac{M}3+1$ \\
$n_\omega$ & $\tfrac{M-2}3$ & $\tfrac{M+2}3$ & $\tfrac{M}3$ \\
$n_{\omega^2}$ & $\tfrac{M+1}3$ & $\tfrac{M-1}3$ & $\tfrac{M}3-1$ \\ \hline
\end{tabular}
\caption{\, $M$:~even, $\alpha = 0$}
\end{subtable}

\vspace{15pt}
\begin{subtable}{\textwidth}\centering
\begin{tabular}{c|ccc}\hline
 & $M = 6m+2$ & $M = 6m+4$ & $M = 6m+6$ \\ \hline
$n_1$ & $\tfrac{M+1}3$ & $\tfrac{M+2}3$ & $\tfrac{M}3$ \\
$n_\omega$ & $\tfrac{M+1}3$ & $\tfrac{M-1}3$ & $\tfrac{M}3$ \\
$n_{\omega^2}$ & $\tfrac{M-2}3$ & $\tfrac{M-1}3$ & $\tfrac{M}3$ \\ \hline
\end{tabular}
\caption{\, $M$:~even, $\alpha = 1/3, 2/3$}
\end{subtable}

\vspace{15pt}
\begin{subtable}{\textwidth}\centering
\begin{tabular}{c|ccc}\hline
 & $M = 6m+1$ & $M = 6m+3$ & $M = 6m+5$ \\ \hline
$n_1$ & $\tfrac{M+2}3$ & $\tfrac{M}3$ & $\tfrac{M+1}3$ \\
$n_\omega$ & $\tfrac{M-1}3$ & $\tfrac{M}3$ & $\tfrac{M+1}3$ \\
$n_{\omega^2}$ & $\tfrac{M-1}3$ & $\tfrac{M}3$ & $\tfrac{M-2}3$ \\ \hline
\end{tabular}
\caption{\, $M$:~odd, $\alpha = 1/6, 5/6$}
\end{subtable}

\vspace{15pt}
\begin{subtable}{\textwidth}\centering
\begin{tabular}{c|ccc}\hline
 & $M = 6m+1$ & $M = 6m+3$ & $M = 6m+5$ \\ \hline
$n_1$ & $\tfrac{M-1}3$ & $\tfrac{M}3+1$ & $\tfrac{M+1}3$ \\
$n_\omega$ & $\tfrac{M+2}3$ & $\tfrac{M}3$ & $\tfrac{M-2}3$ \\
$n_{\omega^2}$ & $\tfrac{M-1}3$ & $\tfrac{M}3-1$ & $\tfrac{M+1}3$ \\ \hline
\end{tabular}
\caption{\, $M$:~odd, $\alpha = 1/2$}
\end{subtable}
}
\caption{The number of independent physical zero modes on $T^2/\mathbb{Z}_3$.}
\label{tab_Z3}
\end{table}

\subsection{$T^2/\mathbb{Z}_4$}
Next, we proceed to $T^2/\mathbb{Z}_4$, and start by considering the $\mathbb{Z}_4$ transformation property for the torus physical states:
\begin{gather}
\hat U_{\mathbb{Z}_4} | \hspace{.5pt} M, j, \alpha, \alpha \rangle_{T^2} = \sum_{k = 0}^{M-1} D_{jk}(\alpha) \hspace{.5pt} | \hspace{.5pt} M, k, \alpha, \alpha \rangle_{T^2},
\end{gather}
where $D_{jk}(\alpha) \equiv D_{jk}(\alpha, \alpha)$ is given in \eqref{D_{jk}}.
Because of $(\hat U_{\mathbb{Z}_4} )^4 = \mathds{1}$, the transformation matrix $D(\alpha)$
gives eigenvalues $1, \omega, \omega^2, \omega^3 \,\, (\omega = i)$.
By an analogous logic, one can see that it leads to
\begin{align}
{\rm tr}\, \bigl( D(\alpha) \bigr) &= n_1 + \omega n_\omega + \omega^2 n_{\omega^2} + \omega^3 n_{\omega^3} \notag \\
&= n_1 - n_{\omega^2} + i(n_\omega - n_{\omega^3}),
\end{align}
where we have used $\omega = i$ and defined $n_{1, \hspace{1pt} \omega, \hspace{1pt} \omega^2, \hspace{1pt} \omega^3}$ as the number of orbifold physical states belonging to $\mathbb{Z}_4$ eigenvalue $\eta = 1, \omega, \omega^2, \omega^3$, respectively.

Note that $\hat U_{\mathbb{Z}_2} \equiv (\hat U_{\mathbb{Z}_4} )^2$ 
behaves as a $\mathbb{Z}_2$ operator and gives eigenvalues $\pm 1$.
Let $| \hspace{.5pt} M, \eta, \alpha, \alpha \rangle_{T^2/\mathbb{Z}_4}$
be a $\mathbb{Z}_{4}$ eigenstate belonging to $\mathbb{Z}_{4}$ eigenvalue $\eta$, i.e.
\begin{gather}
\hat U_{\mathbb{Z}_4} | \hspace{.5pt} M, \eta, \alpha, \alpha \rangle_{T^2/\mathbb{Z}_4} = \eta \hspace{.5pt} | \hspace{.5pt} M, \eta, \alpha, \alpha \rangle_{T^2/\mathbb{Z}_4} \qquad (\eta = 1, \omega, \omega^2, \omega^3)\,.
\end{gather}
Then, this immediately gives
\begin{gather}
(\hat U_{\mathbb{Z}_4})^2 | \hspace{.5pt} M, \omega^\ell, \alpha, \alpha \rangle_{T^2/\mathbb{Z}_4} = 
\begin{cases}
+ | \hspace{.5pt} M, 1, \alpha, \alpha \rangle_{T^2/\mathbb{Z}_4} &\quad (\ell = 0), \\[3pt]
- | \hspace{.5pt} M, \omega, \alpha, \alpha \rangle_{T^2/\mathbb{Z}_4} &\quad (\ell = 1), \\[3pt]
+ | \hspace{.5pt} M, \omega^2, \alpha, \alpha \rangle_{T^2/\mathbb{Z}_4} &\quad (\ell = 2), \\[3pt]
- | \hspace{.5pt} M, \omega^3, \alpha, \alpha \rangle_{T^2/\mathbb{Z}_4} &\quad (\ell = 3).
\end{cases}
\end{gather}
Thus, $(\hat U_{\mathbb{Z}_4})^2$ can be regarded as the $\mathbb{Z}_2$ operator, and the $\mathbb{Z}_4$ orbifold eigenstates for $\ell = 0, 2$ ($\ell = 1, 3$) are $\mathbb{Z}_2$-even (odd) states, respectively.
This is why we can obtain the following relations in terms of $n_{\pm}$ defined 
in Subsection 3.1:
\begin{gather}
n_1 + n_{\omega^2} = n_+ = 
\begin{cases}
\tfrac{M+1}2 &\quad (M = 2m+1, \, \alpha_1 = \alpha_2 = 0), \\[3pt]
\tfrac{M}2 + 1 &\quad (M = 2m+2, \, \alpha_1 = \alpha_2 = 0), \\[3pt]
\tfrac{M-1}2 &\quad (M = 2m + 1, \, \alpha_1 = \alpha_2 = \tfrac12), \\[3pt]
\tfrac{M}2 &\quad (M = 2m+2, \, \alpha_1 = \alpha_2 = \tfrac12),
\end{cases} 
\label{eq3.43}\\
n_\omega + n_{\omega^3} = n_- = 
\begin{cases}
\tfrac{M-1}2 &\quad (M = 2m+1, \, \alpha_1 = \alpha_2 = 0), \\[3pt]
\tfrac{M}2 - 1 &\quad (M = 2m+2, \, \alpha_1 = \alpha_2 = 0), \\[3pt]
\tfrac{M+1}2 &\quad (M = 2m + 1, \, \alpha_1 = \alpha_2 = \tfrac12), \\[3pt]
\tfrac{M}2 &\quad (M = 2m+2, \, \alpha_1 = \alpha_2 = \tfrac12).
\end{cases}
\label{eq3.44}
\end{gather}

\subsubsection*{\underline{$\alpha=0$}}
Using \eqref{D_{jk}} and \eqref{LS1} for $p = M$ and $q = 1$, we can evaluate the trace ${\rm tr}\, \bigl(D(0)\bigr)$ as
\begin{gather}
n_1 - n_{\omega^2} + i(n_\omega - n_{\omega^3}) = {\rm tr}\, \bigl(D(0)\bigr) = 
\begin{cases}
1 &\quad (M = 4m + 1), \\
0 &\quad (M = 4m + 2), \\
i &\quad (M = 4m + 3), \\
1+i &\quad (M = 4m+4).
\end{cases}
\label{eq3.45}
\end{gather}
From \eqref{eq3.43}\,--\,\eqref{eq3.45}, it is straightforward to find
\begin{gather}
n_1 = \frac{M+3}4, \qquad n_\omega = n_{\omega^2} = n_{\omega^3} = \frac{M-1}4 \qquad (M = 4m + 1), \\
n_1 = n_{\omega^2} = \frac{M+2}4, \qquad n_\omega = n_{\omega^3} = \frac{M-2}4 \qquad (M = 4m + 2), \\
n_1 = n_\omega = n_{\omega^2} = \frac{M+1}4, \qquad n_{\omega^3} = \frac{M-3}4 \qquad (M = 4m + 3), \\
n_1 = \frac{M}4+1, \qquad n_\omega = n_{\omega^2} = \frac{M}4, \qquad n_{\omega^3} = \frac{M}4 -1 \qquad (M = 4m+4),
\end{gather}
as summarized in Table \ref{tab_Z4} (a).

\subsubsection*{\underline{$\alpha=1/2$}}
Similarly, using \eqref{D_{jk}} and \eqref{LS3} for $p=M,\ q=2$, and $\nu=1/2$ (with $pq+2q\nu \in 2\mathbb{Z}$ satisfied),
we evaluate the trace ${\rm tr}\, \bigl(D(\tfrac12)\bigr)$ as
\begin{gather}
n_1 - n_{\omega^2} + i(n_\omega - n_{\omega^3}) = {\rm tr}\, \bigl(D(\tfrac12)\bigr) = 
\begin{cases}
i &\quad (M = 4m + 1), \\
1+i &\quad (M = 4m + 2), \\
1 &\quad (M = 4m + 3), \\
0 &\quad (M = 4m + 4).
\end{cases}
\label{eq3.50}
\end{gather}
From \eqref{eq3.43}, \eqref{eq3.44}, and \eqref{eq3.50}, it is straightforward to find
\begin{gather}
n_1 = n_{\omega^2} = n_{\omega^3} = \frac{M-1}4, \qquad n_\omega = \frac{M+3}4 \qquad (M = 4m + 1), \\
n_1 = n_{\omega} = \frac{M+2}4, \qquad n_{\omega^2} = n_{\omega^3} = \frac{M-2}4 \qquad (M = 4m + 2), \\
n_1 = n_\omega = n_{\omega^3} = \frac{M+1}4, \qquad n_{\omega^2} = \frac{M-3}4 \qquad (M = 4m + 3), \\
n_1 = n_\omega = n_{\omega^2} = n_{\omega^3} = \frac{M}4 \qquad (M = 4m+4),
\end{gather}
as summarized in Table \ref{tab_Z4} (b).

\begin{table}[!tb]
\centering
{\tabcolsep = 4mm
\renewcommand{\arraystretch}{1.5}
\begin{subtable}{\textwidth}\centering
\begin{tabular}{c|cccc}\hline
 & $M = 4m+1$ & $M = 4m+2$ & $M = 4m+3$ & $M = 4m+4$ \\ \hline
$n_1$ & $\tfrac{M+3}4$ & $\tfrac{M+2}4$ & $\tfrac{M+1}4$ & $\tfrac{M}4+1$ \\
$n_\omega$ & $\tfrac{M-1}4$ & $\tfrac{M-2}4$ & $\tfrac{M+1}4$ & $\tfrac{M}4$ \\
$n_{\omega^2}$ & $\tfrac{M-1}4$ & $\tfrac{M+2}4$ & $\tfrac{M+1}4$ & $\tfrac{M}4$ \\
$n_{\omega^3}$ & $\tfrac{M-1}4$ & $\tfrac{M-2}4$ & $\tfrac{M-3}4$ & $\tfrac{M}4-1$ \\ \hline
\end{tabular}
\caption{\, $\alpha = 0$}
\end{subtable}

\vspace{15pt}
\begin{subtable}{\textwidth}\centering
\begin{tabular}{c|cccc}\hline
 & $M = 4m+1$ & $M = 4m+2$ & $M = 4m+3$ & $M = 4m+4$ \\ \hline
$n_1$ & $\tfrac{M-1}4$ & $\tfrac{M+2}4$ & $\tfrac{M+1}4$ & $\tfrac{M}4$ \\
$n_\omega$ & $\tfrac{M+3}4$ & $\tfrac{M+2}4$ & $\tfrac{M+1}4$ & $\tfrac{M}4$ \\
$n_{\omega^2}$ & $\tfrac{M-1}4$ & $\tfrac{M-2}4$ & $\tfrac{M-3}4$ & $\tfrac{M}4$ \\
$n_{\omega^3}$ & $\tfrac{M-1}4$ & $\tfrac{M-2}4$ & $\tfrac{M+1}4$ & $\tfrac{M}4$ \\ \hline
\end{tabular}
\caption{\, $\alpha = 1/2$}
\end{subtable}
}
\caption{The number of independent physical zero modes on $T^2/\mathbb{Z}_4$.}
\label{tab_Z4}
\end{table}

\subsection{$T^2/\mathbb{Z}_6$}
Finally, we step into $T^2/\mathbb{Z}_6$.
Although $T^2/\mathbb{Z}_6$ is slightly complicated, the logic here is essentially the same as that in the previous analyses.
Let us start with the $\mathbb{Z}_6$ transformation property of the torus physical states
$| \hspace{.5pt} M, j, \alpha, \alpha \rangle_{T^2}$, i.e.
\begin{gather}
\hat U_{\mathbb{Z}_6} | \hspace{.5pt} M, j, \alpha, \alpha \rangle_{T^2} = \sum_{k = 0}^{M-1} D_{jk}(\alpha) \hspace{.5pt} | \hspace{.5pt} M, k, \alpha, \alpha \rangle_{T^2},
\end{gather}
where $D_{jk}(\alpha) \equiv D_{jk}(\alpha, \alpha)$ is given in \eqref{D_{jk}}.
Because of $(\hat U_{\mathbb{Z}_6} )^6 = \mathds{1}$, the transformation matrix $D_{jk}$ gives eigenvalues $1, \omega, \omega^2, \omega^3, \omega^4, \omega^5 \,\, (\omega = e^{2 \pi i /6})$.
One can again find that this leads to
\begin{align}
{\rm tr}\, \bigl( D(\alpha) \bigr) &= n_1 + \omega n_\omega + \omega^2 n_{\omega^2} + \omega^3 n_{\omega^3} + \omega^4 n_{\omega^4} + \omega^5 n_{\omega^5} \notag \\
&= n_1 - n_{\omega^2} - n_{\omega^3} + n_{\omega^5} + \omega (n_\omega + n_{\omega^2} - n_{\omega^4} - n_{\omega^5}),
\end{align}
where we have used $\omega^2 = \omega - 1$, and defined $n_{1, \hspace{1pt} \omega, \hspace{1pt} \omega^2, \hspace{1pt} \omega^3, \hspace{1pt} \omega^4, \hspace{1pt} \omega^5}$ 
as the number of orbifold $\mathbb{Z}_{6}$ eigenstates belonging to $\mathbb{Z}_6$ eigenvalue 
$\eta = 1, \omega, \omega^2, \omega^3, \omega^4, \omega^5$, respectively.

In the following, we first show that $\hat U_{\mathbb{Z}_2} \equiv (\hat U_{\mathbb{Z}_6} )^3$ ($\hat U_{\mathbb{Z}_3} \equiv (\hat U_{\mathbb{Z}_6} )^2$) behaves as a $\mathbb{Z}_2$ ($\mathbb{Z}_3$) operator and gives eigenvalues $\pm 1$ ($1, e^{2 \pi i/3}, e^{4 \pi i/3}$), as introduced in Subsections 3.1 and 3.2.
Let $| \hspace{.5pt} M, \eta, \alpha, \alpha \rangle_{T^2/\mathbb{Z}_6}$
be a $\mathbb{Z}_{6}$ eigenstate belonging to $\mathbb{Z}_{6}$ eigenvalue
$\eta = \omega^{\ell} \,\, (\ell =0,1,\ldots,5)$, i.e.
\begin{gather}
\hat U_{\mathbb{Z}_6} | \hspace{.5pt} M, \eta, \alpha, \alpha \rangle_{T^2/\mathbb{Z}_6} = \eta \hspace{.5pt} | \hspace{.5pt} M, \eta, \alpha, \alpha \rangle_{T^2/\mathbb{Z}_6}.
\end{gather}
Then, it implies
\begin{gather}
(U_{\mathbb{Z}_6})^3 | \hspace{.5pt} M, \omega^{\ell}, \alpha, \alpha \rangle_{T^2/\mathbb{Z}_6} =
\begin{cases}
+ | \hspace{.5pt} M, 1, \alpha, \alpha \rangle_{T^2/\mathbb{Z}_6} & \quad (\ell = 0), \\
- | \hspace{.5pt} M, \omega, \alpha, \alpha \rangle_{T^2/\mathbb{Z}_6} & \quad (\ell = 1), \\
+ | \hspace{.5pt} M, \omega^2, \alpha, \alpha \rangle_{T^2/\mathbb{Z}_6} & \quad (\ell = 2), \\
- | \hspace{.5pt} M, \omega^3, \alpha, \alpha \rangle_{T^2/\mathbb{Z}_6} & \quad (\ell = 3), \\
+ | \hspace{.5pt} M, \omega^4, \alpha, \alpha \rangle_{T^2/\mathbb{Z}_6} & \quad (\ell = 4), \\
- | \hspace{.5pt} M, \omega^5, \alpha, \alpha \rangle_{T^2/\mathbb{Z}_6} & \quad (\ell = 5).
\end{cases}
\end{gather}
It is confirmed that $(\hat U_{\mathbb{Z}_6})^3$ practically behaves as the $\mathbb{Z}_2$ operator, and the $\mathbb{Z}_6$ orbifold eigenstates belonging to $\mathbb{Z}_{6}$ eigenvalue $\eta = \omega^{\ell}$ for $\ell = 0, 2, 4$ ($\ell = 1, 3, 5$) correspond to $\mathbb{Z}_2$-even (odd) states, respectively.
Now, in terms of $n_{\pm}$ in Subsection 3.1, we reach 
\begin{gather}
n_1 + n_{\omega^2} + n_{\omega^4} = n_+ = 
\begin{cases}
\tfrac{M-1}2 &\quad (M = 2m + 1, \, \alpha_1 = \alpha_2 = \tfrac12), \\[3pt]
\tfrac{M}2 + 1 &\quad (M = 2m + 2, \, \alpha_1 = \alpha_2 = 0),
\end{cases}  \label{eq3.59}\\
n_\omega + n_{\omega^3} + n_{\omega^5} = n_- = 
\begin{cases}
\tfrac{M+1}2 &\quad (M = 2m + 1, \, \alpha_1 = \alpha_2 = \tfrac12), \\[3pt]
\tfrac{M}2 - 1 &\quad (M = 2m + 2, \, \alpha_1 = \alpha_2 = 0).
\end{cases}  \label{eq3.60}
\end{gather}

On the other hand, one can show
\begin{gather}
(\hat{U}_{\mathbb{Z}_6})^2 | \hspace{.5pt} M, \omega^{\ell}, \alpha, \alpha \rangle_{T^2/\mathbb{Z}_6} =
\begin{cases}
+ | \hspace{.5pt} M, 1, \alpha, \alpha \rangle_{T^2/\mathbb{Z}_6} & \quad (\ell = 0), \\
\omega' | \hspace{.5pt} M, \omega, \alpha, \alpha \rangle_{T^2/\mathbb{Z}_6} & \quad (\ell = 1), \\
\omega'^2 | \hspace{.5pt} M, \omega^2, \alpha, \alpha \rangle_{T^2/\mathbb{Z}_6} & \quad (\ell = 2), \\
+ | \hspace{.5pt} M, \omega^3, \alpha, \alpha \rangle_{T^2/\mathbb{Z}_6} & \quad (\ell = 3), \\
\omega' | \hspace{.5pt} M, \omega^4, \alpha, \alpha \rangle_{T^2/\mathbb{Z}_6} & \quad (\ell = 4), \\
\omega'^2 | \hspace{.5pt} M, \omega^5, \alpha, \alpha \rangle_{T^2/\mathbb{Z}_6} & \quad (\ell = 5),
\end{cases}
\end{gather}
with $\omega' \equiv e^{2\pi i/3} \,\, (=\omega^2)$ and then find out that $(\hat U_{\mathbb{Z}_6})^2$ behaves as the $\mathbb{Z}_3$ operator, and the $\mathbb{Z}_6$ orbifold eigenstates belonging to $\mathbb{Z}_{6}$ eigenvalue $\eta = \omega^{\ell}$ for $\ell = 0, 3$ ($\ell = 1, 4$ and $\ell = 2, 5$) correspond to $\mathbb{Z}_{3}$ eigestates belonging to $\mathbb{Z}_3$ eigenvalue $+1$ ($\omega'$ and $\omega'^2$), respectively.
In terms of $n_{1', \hspace{1pt} \omega', \hspace{1pt} \omega'^2}$ in Subsection 3.2,
we reach
\begin{gather}
n_1 + n_{\omega^3} = n_{1'} = 
\begin{cases}
\tfrac{M+1}3 &\quad (M = 6m + 2, \, \alpha = 0), \\[3pt]
\tfrac{M-1}3 &\quad (M = 6m + 4, \, \alpha = 0), \\[3pt]
\tfrac{M}3 + 1 &\quad (M = 6m + 6, \, \alpha = 0),
\end{cases} \label{Z6_alpha=0_1} \\
n_\omega + n_{\omega^4} = n_{\omega'} = 
\begin{cases}
\tfrac{M-2}3 &\quad (M = 6m + 2, \, \alpha = 0), \\[3pt]
\tfrac{M+2}3 &\quad (M = 6m + 4, \, \alpha = 0), \\[3pt]
\tfrac{M}3 &\quad (M = 6m + 6, \, \alpha = 0),
\end{cases} \\
n_{\omega^2} + n_{\omega^5} = n_{\omega'^2} = 
\begin{cases}
\tfrac{M+1}3 & (M = 6m + 2, \, \alpha = 0), \\[3pt]
\tfrac{M-1}3 & (M = 6m + 4, \, \alpha = 0), \\[3pt]
\tfrac{M}3 - 1 & (M = 6m + 6, \, \alpha = 0),
\end{cases} \label{Z6_alpha=0_3}
\end{gather}
and
\begin{gather}
n_1 + n_{\omega^3} = n_{1'} = 
\begin{cases}
\tfrac{M-1}3 &\quad (M = 6m+1, \, \alpha = \tfrac12), \\[3pt]
\tfrac{M}3 + 1 &\quad (M = 6m + 3, \, \alpha = \tfrac12), \\[3pt]
\tfrac{M+1}3 &\quad (M = 6m + 5, \, \alpha = \tfrac12),
\end{cases} \label{Z6_alpha=1/2_1} \\
n_\omega + n_{\omega^4} = n_{\omega'} = 
\begin{cases}
\tfrac{M+2}3 &\quad (M = 6m+1, \, \alpha = \tfrac12), \\[3pt]
\tfrac{M}3 &\quad (M = 6m + 3, \, \alpha = \tfrac12), \\[3pt]
\tfrac{M-2}3 &\quad (M = 6m + 5, \, \alpha = \tfrac12),
\end{cases} \\
n_{\omega^2} + n_{\omega^5} = n_{\omega'^2} = 
\begin{cases}
\tfrac{M-1}3 &\quad (M = 6m+1, \, \alpha = \tfrac12), \\[3pt]
\tfrac{M}3 - 1 &\quad (M = 6m + 3, \, \alpha = \tfrac12), \\[3pt]
\tfrac{M+1}3 &\quad (M = 6m + 5, \, \alpha = \tfrac12).
\end{cases} \label{Z6_alpha=1/2_3}
\end{gather}

\subsubsection*{\underline{$\alpha=0$}}
To evaluate the trace ${\rm tr}\, \bigl( D(0) \bigr)$, 
we need to use \eqref{D_{jk}} and \eqref{LS2} for $p = 1$ and $2q = M = \textrm{even}$.
Then, we find
\begin{gather}
n_1 - n_{\omega^2} - n_{\omega^3} + n_{\omega^5} + \omega (n_\omega + n_{\omega^2} - n_{\omega^4} - n_{\omega^5}) = {\rm tr}\, \bigl( D(0) \bigr) = \omega.
\end{gather}
Comparing this relation with \eqref{eq3.59}, \eqref{eq3.60}, and \eqref{Z6_alpha=0_1}\,--\,\eqref{Z6_alpha=0_3}, we obtain
\begin{gather}
n_1 = n_{\omega^2} = \frac{M+4}6, \quad n_{\omega} = n_{\omega^3} = n_{\omega^4} = n_{\omega^5} = \frac{M-2}6 \qquad (M = 6m + 2), \\
n_1 = n_{\omega} = n_{\omega^2} = n_{\omega^4} = \frac{M+2}6, \quad n_{\omega^3} = n_{\omega^5} = \frac{M-4}6 \qquad (M = 6m + 4), \\
n_1 = \frac{M}6 + 1, \quad n_{\omega} = n_{\omega^2} = n_{\omega^3} = n_{\omega^4} = \frac{M}6, \quad n_{\omega^5} = \frac{M}6 -1 \qquad (M = 6m + 6)\,,
\end{gather}
which are summarized in Table \ref{tab_Z6} (a).

\subsubsection*{\underline{$\alpha=1/2$}}
Using \eqref{D_{jk}} and \eqref{LS3} for $p = M = \textrm{odd}$, $q = 1$, and $\nu = 1/2$
(with $pq + 2q\nu \in 2\mathbb{Z}$ satisfied), one can claim 
\begin{gather}
n_1 - n_{\omega^2} - n_{\omega^3} + n_{\omega^5} + \omega (n_\omega + n_{\omega^2} - n_{\omega^4} - n_{\omega^5}) = {\rm tr}\, \bigl( D(\tfrac12) \bigr) = \omega.
\end{gather}
By comparing this equation with \eqref{eq3.59}, \eqref{eq3.60}, and \eqref{Z6_alpha=1/2_1}\,--\,\eqref{Z6_alpha=1/2_3}, the number of $\mathbb{Z}_{6}$ eigenstates for each $\mathbb{Z}_6$ eigenvalue is given as
\begin{gather}
n_1 = n_{\omega^2} = n_{\omega^3} = n_{\omega^4} = n_{\omega^5} = \frac{M-1}6, \quad n_{\omega} = \frac{M+5}6\qquad (M = 6m + 1), \\
n_1 = n_{\omega} = n_{\omega^3} = \frac{M+3}6, \quad n_{\omega^2} = n_{\omega^4} = n_{\omega^5} = \frac{M-3}6 \qquad (M = 6m + 3), \\
n_1 = n_{\omega} = n_{\omega^2} = n_{\omega^3} = n_{\omega^5} = \frac{M+1}6, \quad n_{\omega^4} = \frac{M-5}6 \qquad (M = 6m + 5)\,,
\end{gather}
which are summarized in Table \ref{tab_Z6} (b).

We should mention that the results given in Tables \ref{tab_Z2}\,--\,\ref{tab_Z6} are consistent with those in \cite{Abe:2013bca, Abe:2014noa},
but the results for the non-vanishing SS twist phases on $T^{2}/\mathbb{Z}_{N} \,\, (N=3,4,6)$ are newly obtained in this paper.
Tables \ref{tab_Z2}\,--\,\ref{tab_Z6} give a complete list for the number of 
the $\mathbb{Z}_{N}$ eigen zero modes on the orbifold $T^{2}/\mathbb{Z}_{N} \,\, (N=2,3,4,6)$, as announced before.

\begin{table}[!tb]
\centering
{\tabcolsep = 5mm
\renewcommand{\arraystretch}{1.5}
\begin{subtable}{\textwidth}\centering
\begin{tabular}{c|ccc}\hline
 & $M = 6m+2$ & $M = 6m+4$ & $M = 6m+6$ \\ \hline
$n_1$ & $\tfrac{M+4}6$ & $\tfrac{M+2}6$ & $\tfrac{M}6+1$ \\
$n_\omega$ & $\tfrac{M-2}6$ & $\tfrac{M+2}6$ & $\tfrac{M}6$ \\
$n_{\omega^2}$ & $\tfrac{M+4}6$ & $\tfrac{M+2}6$ & $\tfrac{M}6$ \\
$n_{\omega^3}$ & $\tfrac{M-2}6$ & $\tfrac{M-4}6$ & $\tfrac{M}6$ \\ 
$n_{\omega^4}$ & $\tfrac{M-2}6$ & $\tfrac{M+2}6$ & $\tfrac{M}6$ \\
$n_{\omega^5}$ & $\tfrac{M-2}6$ & $\tfrac{M-4}6$ & $\tfrac{M}6-1$ \\ \hline
\end{tabular}
\caption{\, $M$:~even, $\alpha = 0$}
\end{subtable}

\vspace{15pt}
\begin{subtable}{\textwidth}\centering
\begin{tabular}{c|ccc}\hline
 & $M = 6m+1$ & $M = 6m+3$ & $M = 6m+5$ \\ \hline
$n_1$ & $\tfrac{M-1}6$ & $\tfrac{M+3}6$ & $\tfrac{M+1}6$ \\
$n_\omega$ & $\tfrac{M+5}6$ & $\tfrac{M+3}6$ & $\tfrac{M+1}6$ \\
$n_{\omega^2}$ & $\tfrac{M-1}6$ & $\tfrac{M-3}6$ & $\tfrac{M+1}6$ \\
$n_{\omega^3}$ & $\tfrac{M-1}6$ & $\tfrac{M+3}6$ & $\tfrac{M+1}6$ \\ 
$n_{\omega^4}$ & $\tfrac{M-1}6$ & $\tfrac{M-3}6$ & $\tfrac{M-5}6$ \\
$n_{\omega^5}$ & $\tfrac{M-1}6$ & $\tfrac{M-3}6$ & $\tfrac{M+1}6$ \\ \hline
\end{tabular}
\caption{\, $M$:~odd, $\alpha = 1/2$}
\end{subtable}
}
\caption{The number of independent physical zero modes on $T^2/\mathbb{Z}_6$.}
\label{tab_Z6}
\end{table}

\section{Analysis of zero points}
\begin{figure}[!t]
\centering
\includegraphics[width=0.8\textwidth]{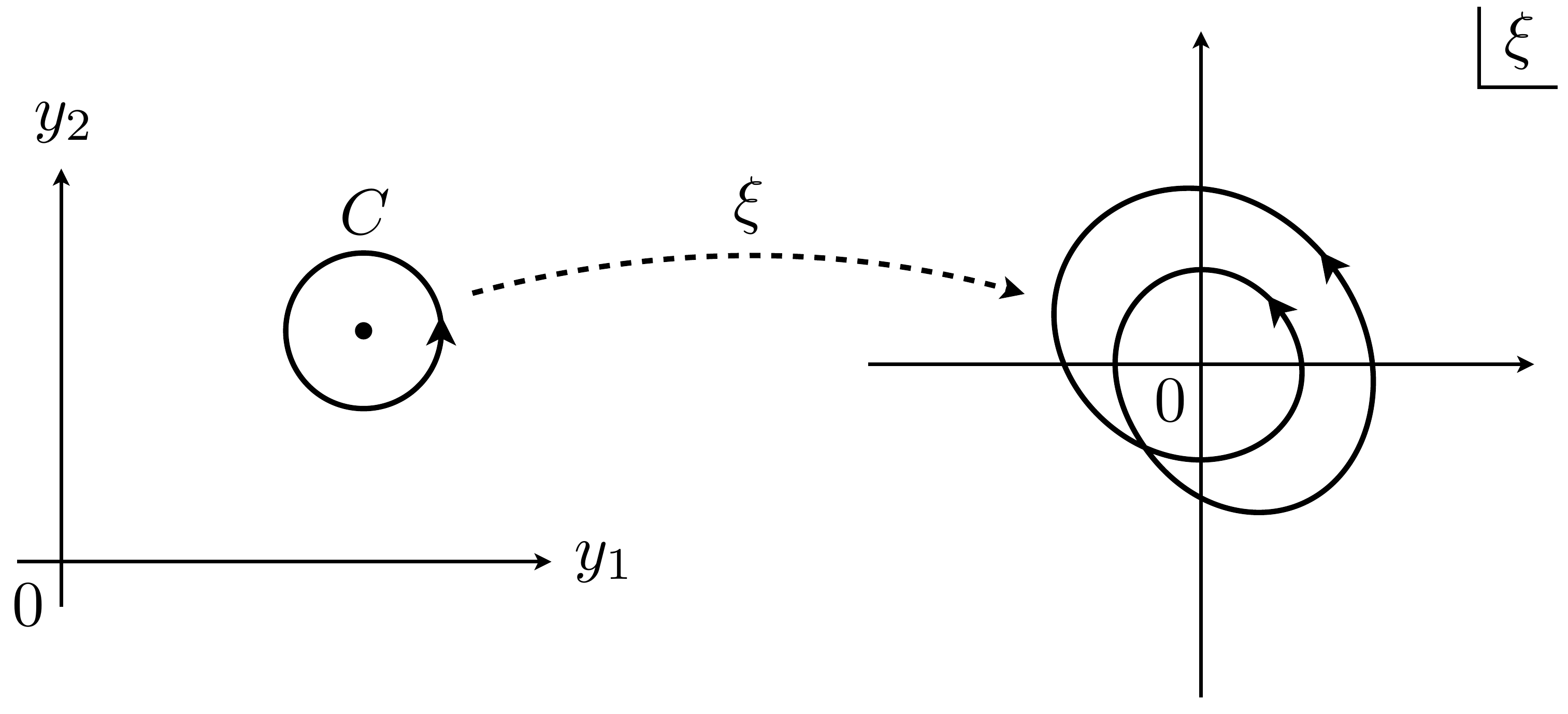}
\caption{Winding number or vortex number that the zero mode wavefunctions yield. In this example, the winding number is $+2$. A black dot in the left figure denotes a zero point of $\xi^j(z)$.}
\label{fredholm}
\end{figure}

We are ready to move on to our main subject.
In the previous section, we have succeeded in obtaining a complete list for the number of the $\mathbb{Z}_{N}$ eigenstates.
It seems hard that all the numbers of the $\mathbb{Z}_{N}$ eigenstates given in Tables \ref{tab_Z2}\,--\,\ref{tab_Z6} can be universally
explained in a simple formula.
That is because those numbers in 
Tables \ref{tab_Z2}\,--\,\ref{tab_Z6} quite complicatedly depend on the flux quanta $M$, the SS twist phase $(\alpha_{1},\alpha_{2})$, and 
the $\mathbb{Z}_{N}$ eigenvalue $\eta = \omega^{\ell}$ ($\ell = 0,1,\ldots,N-1$), 
as well as the $\mathbb{Z}_{N}$
twist $N$.

Surprisingly, it turns out that all the numbers in Tables \ref{tab_Z2}\,--\,\ref{tab_Z6}
can be described by a single zero-mode counting formula
\begin{gather}
n_{\eta} = \frac{M - V_{\eta}}{N} + 1,
\label{countingformula}
\end{gather}
where $n_{\eta}$ is the number of the $\mathbb{Z}_{N}$ eigenstates belonging to the $\mathbb{Z}_{N}$ eigenvalue $\eta$, and
$V_{\eta}$ is the sum of winding numbers at the fixed points of the orbifold $T^{2}/\mathbb{Z}_{N}$.
The formula \eqref{countingformula} is the most important result in this paper.
The details will be given in the following.

Our starting point is the Atiyah-Singer index theorem on the torus $T^2$ 
with magnetic flux background \cite{Witten:1984dg, Green:1987mn, Weinberg:1981eu}, 
\begin{align}
{\rm Ind}\,(i \slashed{D}) &= n_{+} - n_{-} \notag \\
&= \frac{q}{2\pi} \int_{T^2} F = M.
\label{ASindextheorem}
\end{align}
Here $n_{\pm}$ denotes the number of zero modes $\psi_{\pm, \hspace{.5pt} 0}$ \eqref{psipm} on the torus base.
As we have seen, for $M>0$ ($M<0$), only $\psi_{+, \hspace{.5pt} 0}$ ($\psi_{-, \hspace{.5pt} 0}$) possesses $|M|$-fold normalizable zero modes.
That is why we easily see that the index theorem actually holds on the magnetized torus.

There exists another expression of the index theorem, the notion of which is that the index ${\rm Ind}\,(i \slashed{D})$ is exactly equal to the total {\em winding number} 
(or occasionally called {\em vortex number}) \cite{VENUGOPALKRISHNA1972349, Weinberg:1981eu}: 
\begin{align}
{\rm Ind}\,(i \slashed{D}) &= \sum_{i} \frac1{2\pi i} \oint_{C_i} \bm{\nabla} (\log \, \xi^j(z)) \cdot d \bm{\ell} \notag\\
&\equiv \sum_i \chi_i \label{contour_integral} .
\end{align}
This theorem is known as the index theorem for the Fredholm operator (see, for example, \cite{aliprantis2002invitation}).
Here $C_i$ shows an anti-clockwise contour around the zero point  $p_i$ of the torus zero mode $\xi^j (z)$, i.e.
\begin{gather}
\xi^j (z = p_i) = 0.
\end{gather}
The contour integral $\chi_i$ along a contour $C_i$ defines a winding number, i.e. how many times $\xi^j$ wraps around the origin, as illustrated in Figure \ref{fredholm}.
According to the ``residue theorem" in $\xi$ space, the quantity $\chi_i$ is always an integer (see, for example, \cite{krantz1999handbook}).
Note that if there is no zero point inside the contour $C_{i}$, or $p_i$ is not a zero point of $\xi^j$, then $\chi_i$ obviously takes zero due to the ``Cauchy integral formula" in $\xi$ space.

In the following, we will define the winding number $\chi_{i}$ on the fundamental
domain of $T^{2}$ even for the orbifold $T^{2}/\mathbb{Z}_{N}$ and basically evaluate $\chi_{i}$ at the fixed point $z=p_{i}$ on the orbifold.
(See \eqref{fixedpoint} for the fixed points on $T^{2}/\mathbb{Z}_{N}$.)
If one defines the winding number on the fundamental domain of 
the orbifold $T^{2}/\mathbb{Z}_{N}$, instead of $T^{2}$, the sum of the winging number $\chi_{i}$ should be divided
by $N$, i.e. $\sum_{i}\chi_{i}/N$ due to the $1/N$ reduced area and the deficit angles
around the fixed points in comparison with those of the torus.

\begin{figure}[!t]
\centering
\includegraphics[width=0.8\textwidth]{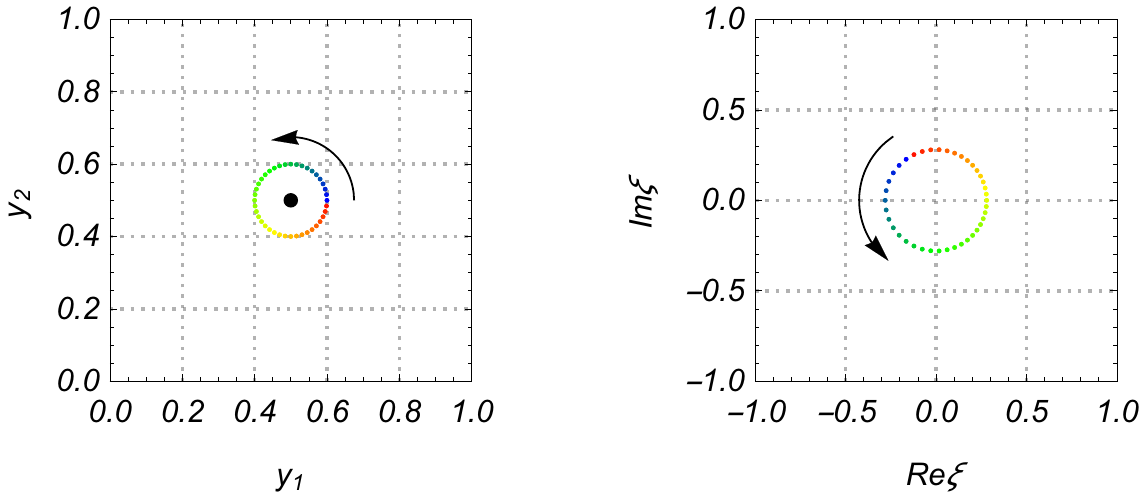}
\caption{Left: a contour with radius $\epsilon = 0.1$ and $\theta\!:\,0 \to 2 \pi$ around a zero point $(y_1, y_2)=(1/2, 1/2)$ depicted by a bullet. Right: its image in zero-mode space.}
\label{vortex}
\end{figure}

Before we tackle the orbifold case, it is instructive to examine \eqref{contour_integral} on the torus.
We start with the zero modes \eqref{toruszeromodes}:
\begin{gather}
\xi^j(z) = \mathcal{N}\,e^{i \pi M z \, {\rm Im}\,z / {\rm Im}\, \tau} \, \vartheta
\begin{bmatrix}
\tfrac{j + \alpha_1}M \\[3pt] -\alpha_2
\end{bmatrix}
(Mz, M\tau).
\end{gather}
Zero points of these zero mode wavefunctions can be obtained as follows.
Setting now $j = 0$ and $\alpha_1 = \alpha_2 = 0$, 
we solve an equation
\begin{gather}
\vartheta
\begin{bmatrix}
0 \\[3pt] 0
\end{bmatrix}
(M (y_1 + \tau y_2), M\tau) = 0.
\end{gather}
The solutions are given by
\begin{gather}
(y_1, y_2) = \left(\frac1{2M}, \hspace{1pt} \frac12 \right), \left(\frac{3}{2M}, \hspace{1pt} \frac12 \right), ..., \left(\frac{2M-1}{2M}, \hspace{1pt} \frac12 \right).
\end{gather}

Let us look at what is happening in $\xi$ space $({\rm Re} \, \xi, {\rm Im}\,\xi)$.
Figure~\ref{vortex} shows an example of the zero mode given by $M=1$ (i.e. $j = 0$) and $\alpha_1 = \alpha_2 = 0$, where an anti-clockwise contour is a circle with radius $\epsilon = 0.1$ on the $T^2$ fundamental domain ($y_1, y_2 \in [0, 1)$) and it gives its image in $\xi$ space.
Then, there is a unique zero point at $(y_1, y_2) = (1/2, 1/2)$, i.e. $z=(1+\tau)/2$.
We define a contour $z = (1 + \tau)/2 + \epsilon \, e^{i \theta}$ around the zero point.
As the contour runs anti-clockwise from blue ($\theta=0$) to red ($\theta = 2\pi$) gradually, the image varies in the same color correspondence in $\xi$ space (see Figure \ref{vortex}).
In this example, it is easy to evaluate
\begin{gather}
\frac1{2\pi i} \oint_{|z - (1 + \tau)/2| = \epsilon} \bm{\nabla} (\log \, \xi^j(z)) \cdot d\bm{\ell} = +1.
\end{gather}
For this observation, we understand that the winding number $\chi_{i}$ \eqref{contour_integral} counts the order of zero at each zero point $p_{i}$, 
based on the ``residue theorem" in $\xi$ space.

Since one can easily confirm that the order of zero is always one, in general, 
for all $M$ and $j = 0, 1, ..., M-1$, we see
\begin{gather}
\frac1{2 \pi i} \oint_{C_i} \bm{\nabla} (\log \, \xi^{j}(z)) \cdot d\bm{\ell} = + 1 \qquad (i=1, 2, ..., M), \\
\Rightarrow \, \sum_{i=1}^M \chi_i = \sum_{i=1}^M \frac1{2 \pi i} \oint_{C_i} \bm{\nabla} (\log \, \xi^{j}(z)) \cdot d\bm{\ell} = + M.
\label{index=M}
\end{gather}
Thus, the winding number $\chi_i$ gives a consistent result with the Atiyah-Singer index theorem \eqref{ASindextheorem}.

It is instructive to show another way to derive \eqref{index=M}.
Along a parallelogram contour $C\!:\,z = 0 \to 1 \to 1 + \tau \to \tau \to 0$, we compute 
\begin{align}
\sum_{p_i \in\,T^2} \chi_i &= \frac1{2\pi i} \oint_C \bm{\nabla} (\log \, \xi^j(z) ) \cdot d\bm{\ell} \notag \\
&= \frac1{2 \pi i} \bigg\{ \int_0^1 dy_1 
\left( \frac{1}{\xi^j(y_1)}\frac{\partial  \xi^j(y_1)}{\partial y_1}  
- \frac{1}{\xi^j(y_1 + \tau)}\frac{\partial  \xi^j(y_1 + \tau)}{\partial y_1} \right) \notag \\
& \hspace{45pt} + \int_0^1 dy_2 
\left( \frac{1}{\xi^j(1 + \tau y_2)}\frac{\partial \xi^j(1 + \tau y_2)}{\partial y_2}  
- \frac{1}{\xi^j(\tau y_2)}\frac{\partial \xi^j(\tau y_2)}{\partial y_2} \right) \bigg\} \notag \\
&= +M, \label{totalwindingnumber}
\end{align}
where we have used the boundary conditions \eqref{BCs} and
\begin{gather}
\frac{\partial \xi^j(z + 1)}{\partial y_2} 
= e^{iq \Lambda_1(z) + 2\pi i\alpha_{1}} 
\left( \frac{iqf}2 + \frac\partial{\partial y_2} \right) \xi^j(z), \\
\frac{\partial \xi^j(z + \tau)}{\partial y_1} 
= e^{iq \Lambda_2(z) + 2\pi i\alpha_{2}} 
\left( -\frac{iqf}2 + \frac\partial{\partial y_1} \right) \xi^j(z).
\end{gather}
Note that the sum of the winding numbers $\chi_i$ along $C$ is determined only by the boundary conditions \eqref{BCs}.

The result \eqref{totalwindingnumber} directly claims that the sum of $\chi_i$, 
namely the index ${\rm Ind}\,(i \slashed{D})$, gives the same outcome even if we take any linear combination 
of the torus zero modes $\xi^j(z) \,\, (j=0,1, ..., M-1)$.
It can be rephrased as 
\begin{gather}
\tilde{\xi}^j(z) \equiv \sum_{k=0}^{M-1} c_{jk} \, \xi^k(z) \quad (c_{jk} \in \mathbb{C}) \notag \\
\Rightarrow \, \frac1{2 \pi i} \oint_{C} \bm{\nabla} (\log \, \tilde{\xi}^j(z)) \cdot d\bm{\ell} = + M
\end{gather}
which follows the fact that $\tilde{\xi}^j(z)$ satisfies the same boundary conditions \eqref{BCs} as those of $\xi^j(z)$.
One has to be careful that the positions of 
the zeros of $\tilde{\xi}^j(z)$ are now different from the original ones $p_i$, in general.

For $j \neq 0$, it is known that $j$ shifts the peak of Gaussian(-like) zero mode wavefunctions along $y_2$-direction \cite{Cremades:2004wa}.
Also, non-zero phases $\alpha_1$ and $\alpha_2$ shift it along $y_2$- and $y_1$-directions, respectively.\footnote{See, for example, \cite{Abe:2013bca}.}
Thus, we find the generic zero points for $\xi^{j}(z)$ \eqref{toruszeromodes} as 
\begin{align}
(y_1, y_2) &= \left(\frac{1/2 + \alpha_2}{M}, \hspace{1pt} \frac12 - \frac{j + \alpha_1}{M} \right), \left(\frac{3/2 + \alpha_2}{M}, \hspace{1pt} \frac12 - \frac{j + \alpha_1}{M} \right), \notag \\
& \hspace{140pt} ..., \left(\frac{(2M - 1)/2 + \alpha_2}{M}, \hspace{1pt} \frac12 - \frac{j + \alpha_1}{M} \right).
\end{align}

The expression \eqref{contour_integral} strongly inspires us to analyze the zero points
of orbifold eigen zero modes. 
It is fair to say that it is hard to derive the index theorem on the orbifolds, due to the singular property of orbifold fixed points.
Then, a primary interest in the past researches has been the number of independent
$\mathbb{Z}_{N}$ eigen zero modes, which depends on the flux quanta $M$, the SS twist phase $(\alpha_{1},\alpha_{2})$, and the $\mathbb{Z}_{N}$ eigenvalue $\eta$. 
However, it has been of less interest to obtain a simple formula counting it in a universal way.
Our primary interest in this paper is to find a single zero-mode counting formula applicable to any pattern.

\subsection{$T^2/\mathbb{Z}_2$}
Hereafter, we omit the degeneracy label $j$ on the torus and orbifolds, unless otherwise stated.
The following discussion basically holds for arbitrary $j$.

In \cite{Buchmuller:2015eya, Buchmuller:2018lkz}, connecting Wilson loops with localized fluxes
at $T^2/\mathbb{Z}_2$ fixed points, the zero points of zero modes at the fixed points have been classified systematically.
In this subsection, we briefly review the zeros on $T^2/\mathbb{Z}_2$.

The starting point here is the $\mathbb{Z}_2$ eigen zero modes in terms of the torus zero modes $\xi(z)$:
\begin{gather}
\xi_{\pm} (z) \equiv \xi(z) \pm \xi(-z).
\end{gather}
Here, the $\mathbb{Z}_{2}$ orbifold eigenstates $\xi_\eta$ are distinguished by the $\mathbb{Z}_{2}$
eigenvalue or the $\mathbb{Z}_2$ parity $\eta = \pm$, i.e.
\begin{gather}
\xi_\eta (-z) = \eta \, \xi_\eta (z).
\end{gather}
It follows from \eqref{BCs} that the eigenfunctions $\xi_\eta$ satisfy  
\begin{align}
\xi_\pm (z) &= \pm \, \xi_\pm(-z), \label{Z2BC_fix1}  \\
\xi_\pm (z+\tfrac12) &= \pm \,e^{iq \Lambda_1(z) + 2 \pi i \alpha_1} \, \xi_\pm (-z + \tfrac12), \\
\xi_\pm (z + \tfrac{\tau}2) &= \pm \, e^{iq \Lambda_2(z) + 2\pi i \alpha_2} \, \xi_\pm (-z + \tfrac{\tau}2), \\
\xi_\pm (z + \tfrac12 + \tfrac{\tau}2) &= \pm \, e^{iq \Lambda_1(z) + iq \Lambda_2(z) + 2\pi i (M/2 + \alpha_1 + \alpha_2)} \, \xi_\pm (-z + \tfrac12 + \tfrac{\tau}2).
\label{Z2BC_fix4}
\end{align}

By plugging $z=0$ into these four relations, we find that the $\mathbb{Z}_{2}$ 
eigenfunctions $\xi_{\pm}(z)$ take zeros at the following fixed points:
\begin{gather}
\xi_- (0) = 0,  \label{eq4.22}\\
\begin{cases}
\xi_- (\tfrac12) = 0 &\quad \text{for}~\alpha_1 = 0, \\[3pt]
\xi_+ (\tfrac12) = 0 &\quad \text{for}~\alpha_1 = \tfrac12,
\end{cases}  \label{eq4.23}\\
\begin{cases}
\xi_- (\tfrac{\tau}2) = 0 &\quad \text{for}~\alpha_2 = 0, \\[3pt]
\xi_+ (\tfrac{\tau}2) = 0 &\quad \text{for}~\alpha_2 = \tfrac12,
\end{cases}  \label{eq4.24}\\
\begin{cases}
\xi_- (\tfrac12 + \tfrac{\tau}2) = 0 &\quad \text{for}~M=2m, \alpha_1 + \alpha_2 = 0, 1~~\text{or}~~M=2m+1, \alpha_1 + \alpha_2 = \tfrac12, \\[3pt]
\xi_+ (\tfrac12 + \tfrac{\tau}2) = 0 &\quad \text{for}~M=2m, \alpha_1 + \alpha_2 = \tfrac12~~\text{or}~~M=2m+1, \alpha_1 + \alpha_2 = 0, 1.  \label{eq4.25}
\end{cases}
\end{gather}
It follows from \eqref{Z2BC_fix1}\,--\,\eqref{Z2BC_fix4} that we can compute
the winding numbers $\chi_i \,\, (i = 1,2,3,4)$ around the fixed points
\begin{gather}
p_1 = 0, \qquad p_2 = 1/2, \qquad p_3 = \tau/2, \qquad p_4 = (1 + \tau)/2
\end{gather}
with a sufficiently small contour $C_{i}$ around $p_i$ for each $i$.
These results are summarized in Table \ref{Z2tab}.

\begin{table}[!t]
\centering
{\tabcolsep = 4mm
\renewcommand{\arraystretch}{1.2}
\scalebox{0.85}{
\begin{tabular}{ccc|cccc|c|c} \hline
flux & parity & twist & \multicolumn{4}{c|}{winding number} &  total & $n_{\eta}$\\
$M$ & $\eta$ & $(\alpha_1, \alpha_2)$ & $\chi_1$ & $\chi_2$ & $\chi_3$ & $\chi_4$ & $V_{\eta}=\sum_i \chi_i$  & $(M-V_{\eta})/2+1$\\ \hline
$2m+1$ & $+1$ & $(0,0)$ & $0$ & $0$ & $0$ & $+1$ & $+1$ & $(M+1)/2$\\
&& $(\tfrac12, 0)$ & $0$ & $+1$ & $0$ & $0$ & $+1$ & $(M+1)/2$ \\
&& $(0, \tfrac12)$ & $0$ & $0$ & $+1$ & $0$ & $+1$ & $(M+1)/2$ \\
&& $(\tfrac12, \tfrac12)$ & $0$ & $+1$ & $+1$ & $+1$ & $+3$ & $(M-1)/2$ \\ \cline{2-9}
& $-1$ & $(0,0)$ & $+1$ & $+1$ &$+1$ & $0$ & $+3$ & $(M-1)/2$ \\
&& $(\tfrac12,0)$ & $+1$ & $0$ & $+1$ & $+1$ & $+3$ & $(M-1)/2$ \\
&& $(0, \tfrac12)$ & $+1$ & $+1$ & $0$ & $+1$ & $+3$ & $(M-1)/2$ \\
&& $(\tfrac12, \tfrac12)$ & $+1$ & $0$ & $0$ & $0$ & $+1$ & $(M+1)/2$ \\ \hline
$2m+2$ & $+1$ & $(0,0)$ & $0$ & $0$ & $0$ & $0$ & $0$ & $M/2+1$ \\
&& $(\tfrac12, 0)$ & $0$ & $+1$ & $0$ & $+1$ & $+2$ & $M/2$ \\
&& $(0, \tfrac12)$ & $0$ & $0$ & $+1$ & $+1$ & $+2$ & $M/2$ \\
&& $(\tfrac12, \tfrac12)$ & $0$ & $+1$ & $+1$ & $0$ & $+2$ & $M/2$ \\ \cline{2-9}
& $-1$ & $(0,0)$ & $+1$ & $+1$ & $+1$ & $+1$ & $+4$ & $M/2-1$ \\
&& $(\tfrac12, 0)$ & $+1$ & $0$ & $+1$ & $0$ & $+2$ & $M/2$ \\
&& $(0, \tfrac12)$ & $+1$ & $+1$ & $0$ & $0$ & $+2$ & $M/2$ \\
&& $(\tfrac12, \tfrac12)$ & $+1$ & $0$ & $0$ & $+1$ & $+2$ & $M/2$ \\ \hline
\end{tabular}
}
}
\caption{The winding number $\chi_{i}$ at the fixed point $p_{i}\ (i=1,2,3,4)$ (see also \cite{Buchmuller:2018lkz}).
All the values of $(M-V_{\eta})/2+1$ exactly agree with the numbers $n_{\eta}$ of the
$\mathbb{Z}_{2}$ physical zero modes given in Table \ref{tab_Z2}.}
\label{Z2tab}
\end{table}

One should notice the difference between zeros at the fixed points
and those on the bulk.
To this end, let us consider an example of three flux quanta $M=3$ and
a trivial twist phase $\alpha_{1} = \alpha_{2} = 0$.
Then, we have two $\eta = +1$ eigen zero modes on $T^2/\mathbb{Z}_2$, 
say $\xi_+^0(z)$ and $\xi_+^1(z)$ (see Table \ref{tab_Z2} (a)).
From \eqref{eq4.25}, they are vanishing at the fixed point $z = p_4$, i.e.
\begin{gather}
\xi_+^0 (p_4) = \xi_+^1 (p_4) = 0.
\label{eg_removable}
\end{gather}
Note that $\xi_{+}^{0}(z)$ and $\xi_{+}^{1}(z)$ take non-zero values at the other fixed points $z=p_{1}, p_{2}, p_{3}$
(see \eqref{eq4.22}\,--\,\eqref{eq4.24}).

There are additional two zero points on the bulk of $ T^{2}$ 
for each $\xi_+^0$ and $\xi_+^1$,
because each of $\xi^{0}_{+}$ and $\xi^{1}_{+}$ should possess three zero points.
In general, once we take their linear combination, we need to search for new zero points.
In other words, even if we find two zeros $p^0$ ($p^1$) on the bulk such that 
$\xi_+^0(p^{0}) = 0$ ($\xi_+^1(p^{1}) = 0$), a linear combination $c \, \xi_+^0 + c' \, \xi_+^1 \,\, (c, c' \in \mathbb{C})$ does not always vanish at both $p^0$ and $p^1$.
Thus, such an observation inspires us to call them {\em removable zeros}, because their positions of zeros are changeable by taking some linear combination of
$\xi^{0}_{+}$ and $\xi^{1}_{+}$.

On the other hand, because of \eqref{eg_removable}, we easily see $c \, \xi_+^0(p_4) + c' \, \xi_+^1(p_4) = 0$ for arbitrary $c, c'$.
The zero at $p_4$ cannot be removed by taking any linear combination.
Hence, it is reasonable that zeros at the orbifold fixed points are called {\em unremovable zeros}.\footnote{
In the context of string theory on orbifolds \cite{Green:1987mn}, 
unremovable zeros correspond to
twisted strings, which cannot escape from fixed points. 
}
It also implies that there is no need to take removable zeros seriously, 
since the positions of removable zeros are no longer important.

\subsection{$T^2/\mathbb{Z}_3$}
For $\tau = \omega = e^{2 \pi i /3}$, we begin with $\mathbb{Z}_{3}$ eigen zero modes,
\begin{gather}
\xi_\eta (z) = \sum_{\ell = 0}^2 \bar \eta^\ell \, \xi(\omega^\ell z) \qquad (\eta = 1, \hspace{1pt} \omega, \hspace{1pt} \omega^2)
\end{gather}
which belong to the $\mathbb{Z}_3$ eigenvalue $\eta = \omega^k$:
\begin{gather}
\xi_{\omega^{k}} (\omega z) = {\omega^{k}} \, \xi_{\omega^{k}} (z)
 \qquad (k=0,1,2).  \label{eq4.29}
\end{gather}
In analogy to the previous subsection, we can straightforwardly show
\begin{align}
\xi_{\omega^{k}} (\omega z + \tfrac23 + \tfrac{\tau}3) 
&= e^{- iq \Lambda_1(z) - iq \Lambda_2(z) - 2 \pi i (M/3 + 2\alpha - k/3 )} \, 
\xi_{\omega^{k}} (z + \tfrac23 + \tfrac{\tau}3 ),  \label{eq4.30}\\
\xi_{\omega^{k}} (\omega z + \tfrac13 + \tfrac{2\tau}3) 
&= e^{iq \Lambda_1(\omega z) + iq \Lambda_2(\omega z) + 2 \pi i (2M/3 + 2\alpha + k/3 )} 
\, \xi_{\omega^{k}} (z + \tfrac13 + \tfrac{2\tau}3 ).  \label{eq4.31}
\end{align}
Ignoring the terms related to $\Lambda_1 (z)$ and $\Lambda_2(z)$
for infinitesimally small $|z|$, the relations \eqref{eq4.29}\,--\,\eqref{eq4.31} 
reduce to
\begin{align}
\xi_{\omega^{k}} (\omega z) 
&= {\omega^{k}} \, \xi_{\omega^{k}}(z), \label{eq4.32}\\
\xi_{\omega^{k}} (\omega z + \tfrac23 + \tfrac{\tau}3) 
&= e^{- 2 \pi i (M/3 + 2\alpha - k/3 )}
\, \xi_{\omega^{k}} (z + \tfrac23 + \tfrac{\tau}3 ),  \label{eq4.33}\\
\xi_{\omega^{k}} (\omega z + \tfrac13 + \tfrac{2\tau}3) 
&= e^{2 \pi i (2M/3 + 2\alpha + k/3 )} 
\, \xi_{\omega^{k}} (z + \tfrac13 + \tfrac{2\tau}3 ) \qquad (k=0,1,2),  \label{eq4.34}
\end{align}

The above relations tell the phase shifts to the $\mathbb{Z}_{3}$
eigen zero modes $\xi_{\omega^{k}}$ when rotated by $2\pi/3$ around the fixed points.
To evaluate the winding numbers $\chi_i$ at the fixed points $p_{i}\ (i=1,2,3)$,
all we should do is to utilize the above relations three times repeatedly.
Then, taking $C_{i}$ to be a sufficiently small contour around $p_i$ for each $i$, we obtain
\begin{gather}
\chi_1 = k ~~ {\rm mod}~3, \label{eq4.35} \\
\chi_{2} = -M -6\alpha + k ~~ {\rm mod}~3, \label{eq4.36}\\
\chi_{3} = 2M + 6\alpha + k ~~ {\rm mod}~3, \label{eq4.37}
\end{gather}
for $\xi_{\omega^{k}}$ ($k=0,1,2$).
Here, $\chi_{i} \,\, (i=1,2,3)$ has been defined around the three 
$\mathbb{Z}_3$ orbifold fixed points:
\begin{gather}
p_1 = 0, \qquad p_2 = (2 + \tau)/3, \qquad p_3 = (1 + 2\tau)/3.
\end{gather}
The results in this subsection are summarized in Table \ref{Z3tab}.

We should make two comments on the winding number $\chi_i \,\, (i=1,2,3)$.
As one can see from \eqref{eq4.35}\,--\,\eqref{eq4.37}, $\chi_i$ at the fixed point $p_i$ is less than three, i.e. $\chi_i = 0,1,2$.
If an orbifold zero mode wavefunction gives a winding number larger than or equal to three, it accidentally contains some contribution from removable zeros.
In other words, some removable zeros accidentally coincide unremovable zeros at the fixed points and then enhance the value of $\chi_i$.
By taking an appropriate linear combination of orbifold zero modes, we can find that the winding number is less than three.

The second comment is that we here consider the fundamental domain of $T^{2}$ 
but not that of $T^{2}/\mathbb{Z}_{3}$ in order to define the winding number $\chi_{i}$.
We have defined the winding number $\chi_{i}$ in \eqref{contour_integral},
where the contour $C_{i}$ is taken to be a circle encircling 
the fixed point $p_{i}$.
If the winding number $\chi_{i}$ is defined on the fundamental domain of the $T^{2}/\mathbb{Z}_{3}$ orbifold,
it should be divided by $N=3$ due to deficit angles around the fixed points.

\begin{table}[!ht]
\centering
{\tabcolsep = 4.5mm
\renewcommand{\arraystretch}{1.2}
\scalebox{0.85}{
\begin{tabular}{ccc|ccc|c|c} \hline
flux & parity & twist & \multicolumn{3}{c|}{winding number} & total & $n_{\eta}$\\
$M$ & $\eta$ & $\alpha$ & $\chi_1$ & $\chi_2$ & $\chi_3$ & $V_{\eta}=\sum_i \chi_i$ &
$(M-V_{\eta})/3+1$\\ \hline
$6m+1$ & $1$ & $1/6$ & $0$ & $+1$ & $0$ & $+1$ & $(M+2)/3$ \\
&& $1/2$ & $0$ & $+2$ & $+2$ & $+4$  & $(M-1)/3$ \\
&& $5/6$ & $0$ & $0$ & $+1$ & $+1$ & $(M+2)/3$  \\ \cline{2-8}
& $\omega$ & $1/6$ & $+1$ & $+2$ & $+1$ & $+4$ & $(M-1)/3$  \\
&& $1/2$ & $+1$ & $0$ & $0$ & $+1$ & $(M+2)/3$  \\
&& $5/6$ & $+1$ & $+1$ & $+2$ & $+4$ & $(M-1)/3$  \\ \cline{2-8}
& $\omega^2$ & $1/6$ & $+2$ & $0$ & $+2$ & $+4$ & $(M-1)/3$  \\
&& $1/2$ & $+2$ & $+1$ & $+1$ & $+4$ & $(M-1)/3$  \\
&& $5/6$ & $+2$ & $+2$ & $0$ & $+4$ & $(M-1)/3$  \\ \hline
$6m+2$ & $1$ & $0$ & $0$ & $+1$ & $+1$ & $+2$ & $(M+1)/3$  \\
&& $1/3$ & $0$ & $+2$ & $0$ & $+2$ & $(M+1)/3$  \\
&& $2/3$ & $0$ & $0$ & $+2$ & $+2$ & $(M+1)/3$  \\ \cline{2-8}
& $\omega$ & $0$ & $+1$ & $+2$ & $+2$ & $+5$ & $(M-2)/3$  \\
&& $1/3$ & $+1$ & $0$ & $+1$ & $+2$ & $(M+1)/3$  \\
&& $2/3$ & $+1$ & $+1$ & $0$ & $+2$ & $(M+1)/3$  \\ \cline{2-8}
& $\omega^2$ & $0$ & $+2$ & $0$ & $0$ & $+2$ & $(M+1)/3$  \\
&& $1/3$ & $+2$ & $+1$ & $+2$ & $+5$ & $(M-2)/3$  \\
&& $2/3$ & $+2$ & $+2$ & $+1$ & $+5$ & $(M-2)/3$  \\ \hline
$6m+3$ & $1$ & $1/6$ & $0$ & $+2$ & $+1$ & $+3$ & $M/3$  \\
&& $1/2$ & $0$ & $0$ & $0$ & $0$ & $M/3+1$  \\
&& $5/6$ & $0$ & $+1$ & $+2$ & $+3$ & $M/3$  \\ \cline{2-8}
& $\omega$ & $1/6$ & $+1$ & $0$ & $+2$ & $+3$ & $M/3$  \\
&& $1/2$ & $+1$ & $+1$ & $+1$ & $+3$ & $M/3$  \\
&& $5/6$ & $+1$ & $+2$ & $0$ & $+3$ & $M/3$  \\ \cline{2-8}
& $\omega^2$ & $1/6$ & $+2$ & $+1$ & $0$ & $+3$ & $M/3$  \\
&& $1/2$ & $+2$ & $+2$ & $+2$ & $+6$ & $M/3-1$  \\
&& $5/6$ & $+2$ & $0$ & $+1$ & $+3$ & $M/3$  \\ \hline
\end{tabular}
}
}
\caption{
The winding number $\chi_{i}$ at the fixed point $p_{i}\ (i=1,2,3)$.
All the values of $(M-V_{\eta})/3+1$ exactly agree with the numbers $n_{\eta}$ of the
$\mathbb{Z}_{3}$ physical zero modes given in Table \ref{tab_Z3}.}
\label{Z3tab}
\end{table}

\addtocounter{table}{-1}
\begin{table}[!ht]
\centering
{\tabcolsep = 4.5mm
\renewcommand{\arraystretch}{1.2}
\scalebox{0.85}{
\begin{tabular}{ccc|ccc|c|c} \hline
flux & parity & twist & \multicolumn{3}{c|}{winding number} & total & $n_{\eta}$\\
$M$ & $\eta$ & $\alpha$ & $\chi_1$ & $\chi_2$ & $\chi_3$ & $V_{\eta}=\sum_i \chi_i$ &
$(M-V_{\eta})/3+1$\\ \hline
$6m+4$ & $1$ & $0$ & $0$ & $+2$ & $+2$ & $+4$ & $(M-1)/3$ \\
&& $1/3$ & $0$ & $0$ & $+1$ & $+1$ & $(M+2)/3$ \\
&& $2/3$ & $0$ & $+1$ & $0$ & $+1$ & $(M+2)/3$ \\ \cline{2-8}
& $\omega$ & $0$ & $+1$ & $0$ & $0$ & $+1$ & $(M+2)/3$ \\
&& $1/3$ & $+1$ & $+1$ & $+2$ & $+4$ & $(M-1)/3$ \\
&& $2/3$ & $+1$ & $+2$ & $+1$ & $+4$ & $(M-1)/3$ \\ \cline{2-8}
& $\omega^2$ & $0$ & $+2$ & $+1$ & $+1$ & $+4$ & $(M-1)/3$ \\
&& $1/3$ & $+2$ & $+2$ & $0$ & $+4$ & $(M-1)/3$ \\
&& $2/3$ & $+2$ & $0$ & $+2$ & $+4$ & $(M-1)/3$ \\ \hline
$6m+5$ & $1$ & $1/6$ & $0$ & $0$ & $+2$ & $+2$ & $(M+1)/3$ \\
&& $1/2$ & $0$ & $+1$ & $+1$ & $+2$ & $(M+1)/3$ \\
&& $5/6$ & $0$ & $+2$ & $0$ & $+2$ & $(M+1)/3$ \\ \cline{2-8}
& $\omega$ & $1/6$ & $+1$ & $+1$ & $0$ & $+2$ & $(M+1)/3$ \\
&& $1/2$ & $+1$ & $+2$ & $+2$ & $+5$ & $(M-2)/3$ \\
&& $5/6$ & $+1$ & $0$ & $+1$ & $+2$ & $(M+1)/3$ \\ \cline{2-8}
& $\omega^2$ & $1/6$ & $+2$ & $+2$ & $+1$ & $+5$ & $(M-2)/3$ \\
&& $1/2$ & $+2$ & $0$ & $0$ & $+2$ & $(M+1)/3$ \\
&& $5/6$ & $+2$ & $+1$ & $+2$ & $+5$ & $(M-2)/3$ \\ \hline
$6m+6$ & $1$ & $0$ & $0$ & $0$ & $0$ & $0$ & $M/3+1$ \\
&& $1/3$ & $0$ & $+1$ & $+2$ & $+3$ & $M/3$ \\
&& $2/3$ & $0$ & $+2$ & $+1$ & $+3$ & $M/3$ \\ \cline{2-8}
& $\omega$ & $0$ & $+1$ & $+1$ & $+1$ & $+3$ & $M/3$ \\
&& $1/3$ & $+1$ & $+2$ & $0$ & $+3$ & $M/3$ \\
&& $2/3$ & $+1$ & $0$ & $+2$ & $+3$ & $M/3$ \\ \cline{2-8}
& $\omega^2$ & $0$ & $+2$ & $+2$ & $+2$ & $+6$ & $M/3-1$ \\
&& $1/3$ & $+2$ & $0$ & $+1$ & $+3$ & $M/3$ \\
&& $2/3$ & $+2$ & $+1$ & $0$ & $+3$ & $M/3$ \\ \hline
\end{tabular}
}
}
\caption{
(Continued.) The winding number $\chi_{i}$ at the fixed point $p_{i}\ (i=1,2,3)$.
All the values of $(M-V_{\eta})/3+1$ exactly agree with the numbers $n_{\eta}$ of the
$\mathbb{Z}_{3}$ physical zero modes given in Table \ref{tab_Z3}.}
\end{table}

\subsection{$T^2/\mathbb{Z}_4$}
As previously noted, 
there are two fixed points under the $\mathbb{Z}_4$ identification $z \sim i z$,
i.e.
\begin{gather}
z = 0 \,\, (\equiv p_1), \quad (1 + i)/2 \,\, (\equiv p_2). \label{Z4fp}
\end{gather}
Since the $\mathbb{Z}_4$ group includes $\mathbb{Z}_{2}$ as its subgroup, 
there are additionally two ``$\mathbb{Z}_2$ fixed points" that are not invariant 
under the $\mathbb{Z}_4$ rotation, but invariant under such a partial $\mathbb{Z}_2$ transformation ($z \to -z$) 
up to torus lattice shifts.
The two $\mathbb{Z}_2$ fixed points are given by
\begin{gather}
z = 1/2 \,\, (\equiv p_3), \quad i/2 \,\, (\equiv p_4). \label{Z2fp}
\end{gather}
As we shall see later, the winding numbers not only at the $\mathbb{Z}_{4}$ fixed points \eqref{Z4fp}
but also at the $\mathbb{Z}_{2}$ fixed points \eqref{Z2fp} contribute to
the zero-mode counting formula \eqref{countingformula} as unremovable zeros.

For $\tau = \omega = i$, we start with $\mathbb{Z}_{4}$ eigen zero modes, given as
\begin{gather}
\xi_\eta (z) = \sum_{\ell=0}^3 \bar \eta^\ell \xi(\omega^\ell z) 
\qquad (\eta = 1, \hspace{1pt} \omega, \hspace{1pt} \omega^2, \hspace{1pt} \omega^3)
\end{gather}
which belong to the $\mathbb{Z}_{4}$ eigenvalue $\eta = \omega^{k}$:
\begin{gather}
\xi_{\omega^{k}} (\omega z) = {\omega^{k}} \xi_{\omega^{k}} (z)
\qquad (k=0,1,2,3).
\end{gather}
Around the $\mathbb{Z}_4$ fixed point $p_2$ and 
the $\mathbb{Z}_2$ ones $p_{3, \hspace{.5pt} 4}$, we can derive the relations
\begin{align}
\xi_{\omega^k} (\omega z + \tfrac12 + \tfrac{\tau}2) &= e^{-iq \Lambda_2(z) - 2 \pi i (-M/4 + \alpha - k/4)} \, \xi_{\omega^k} (z + \tfrac12 + \tfrac{\tau}2), \\
\xi_{\omega^k} (\omega^2 z + \tfrac12) &= e^{-iq \Lambda_1(z) - 2 \pi i (\alpha -k/2 )} \, \xi_{\omega^k} (z + \tfrac12), \\
\xi_{\omega^k} (\omega^2 z + \tfrac{\tau}2) &= e^{-iq \Lambda_2(z) - 2\pi i (\alpha - k/2)} \, \xi_{\omega^k} (z + \tfrac{\tau}2) \qquad (k=0,1,2,3).
\end{align}
Ignoring the terms related to $\Lambda_1(z)$ and $\Lambda_2(z)$ for infinitesimally small $|z|$, we find
\begin{align}
\xi_{\omega^k}(\omega z) &= \omega^k \xi_{\omega^k}(z), \\
\xi_{\omega^k} (\omega z + \tfrac12 + \tfrac{\tau}2) &= e^{- 2 \pi i (-M/4 + \alpha - k/4)} \, \xi_{\omega^k} (z + \tfrac12 + \tfrac{\tau}2), \\ 
\xi_{\omega^k} (\omega^2 z + \tfrac12) &= e^{- 2 \pi i (\alpha -k/2 )} \, \xi_{\omega^k} (z + \tfrac12), \\
\xi_{\omega^k} (\omega^2 z + \tfrac{\tau}2) &= e^{- 2\pi i (\alpha - k/2)} \, \xi_{\omega^k} (z + \tfrac{\tau}2) \qquad (k =0,1,2,3).
\end{align}

Suppose that $C_i$ is a sufficiently small contour around the fixed point $p_i$ for each $i$.
Our results of interest are given as
\begin{gather}
\chi_1 = k ~~ {\rm mod}~4, \label{eq4.50}\\
\chi_2 = M - 4\alpha + k ~~ {\rm mod}~4, \label{eq4.51} \\
\chi_3 = \chi_4 = -2 \alpha + k ~~ {\rm mod}~2, \label{eq4.52}
\end{gather}
for $\xi_{\omega^k} \,\, (k = 0, 1, 2, 3)$.
Here, the winding number $\chi_{i} \,\, (i=1,2,3,4)$ for $\xi_{\omega^{k}}$
has been defined around the fixed point $p_{i} \,\, (i=1,2,3,4)$, respectively.
The results in this subsection are summarized in Table \ref{Z4tab}.
An interesting observation is that although the ``$\mathbb{Z}_2$ fixed points" are not
invariant under the $\mathbb{Z}_{4}$ identification, zero points at the ``$\mathbb{Z}_2$ fixed points" 
appear as unremovable zeros, and their contribution is indispensable to guarantee
the counting formula \eqref{countingformula}.

We comment on the winding number $\chi_i \,\, (i=1,2,3,4)$.
As one can see from \eqref{eq4.50}\,--\,\eqref{eq4.52}, $\chi_{1, \hspace{.5pt} 2}$ ($\chi_{3, \hspace{.5pt} 4}$) at the fixed point $p_{1, \hspace{.5pt} 2}$ ($p_{3, \hspace{.5pt} 4}$) are less than four (two), i.e. $\chi_{1, \hspace{.5pt} 2} = 0,1,2,3$ ($\chi_{3, \hspace{.5pt} 4} = 0,1 $).
If an orbifold zero mode wavefunction gives a winding number at $p_{1, \hspace{.5pt} 2}$ ($p_{3, \hspace{.5pt} 4}$) larger than or equal to four (two), it accidentally contains some contribution from removable zeros.
In other words, some removable zeros accidentally coincide unremovable zeros at the fixed points and then enhance the value of $\chi_i$.
By taking an appropriate linear combination of orbifold zero modes, we can find that the winding number is less than four or two.

\begin{table}[!ht]
\centering
{\tabcolsep = 4.5mm
\renewcommand{\arraystretch}{1.2}
\scalebox{0.85}{
\begin{tabular}{ccc|cc:cc|c|c} \hline
flux & parity & twist & \multicolumn{4}{c|}{winding number} & total & $n_{\eta}$ \\
$M$ & $\eta$ & $\alpha$ & $\chi_1$ & $\chi_2$ & $\chi_3$ & $\chi_4$ & $V_{\eta}=\sum_i \chi_i$
& $(M-V_{\eta})/4+1$ \\ \hline
$4m+1$ & $1$ & $0$ & $0$ & $+1$ & $0$ & $0$ & $+1$ & $(M+3)/4$ \\
&& $1/2$ & $0$ & $+3$ & $+1$ & $+1$ & $+5$ & $(M-1)/4$ \\ \cline{2-9}
& $i$ & $0$ & $+1$ & $+2$ & $+1$ & $+1$ & $+5$ & $(M-1)/4$ \\
&& $1/2$ & $+1$ & $0$ & $0$ & $0$ & $+1$ & $(M+3)/4$ \\ \cline{2-9}
& $-1$ & $0$ & $+2$ & $+3$ & $0$ & $0$ & $+5$ & $(M-1)/4$ \\
&& $1/2$ & $+2$ & $+1$ & $+1$ & $+1$ & $+5$ & $(M-1)/4$ \\ \cline{2-9}
& $-i$ & $0$ & $+3$ & $0$ & $+1$ & $+1$ & $+5$ & $(M-1)/4$ \\
&& $1/2$ & $+3$ & $+2$ & $0$ & $0$ & $+5$ & $(M-1)/4$ \\ \hline
$4m+2$ & $1$ & $0$ & $0$ & $+2$ & $0$ & $0$ & $+2$ & $(M+2)/4$ \\
&& $1/2$ & $0$ & $0$ & $+1$ & $+1$ & $+2$ & $(M+2)/4$ \\ \cline{2-9}
& $i$ & $0$ & $+1$ & $+3$ & $+1$ & $+1$ & $+6$ & $(M-2)/4$ \\
&& $1/2$ & $+1$ & $+1$ & $0$ & $0$ & $+2$ & $(M+2)/4$ \\ \cline{2-9}
& $-1$ & $0$ & $+2$ & $0$ & $0$ & $0$ & $+2$ & $(M+2)/4$ \\
&& $1/2$ & $+2$ & $+2$ & $+1$ & $+1$ & $+6$ & $(M-2)/4$ \\ \cline{2-9}
& $-i$ & $0$ & $+3$ & $+1$ & $+1$ & $+1$ & $+6$ & $(M-2)/4$ \\
&& $1/2$ & $+3$ & $+3$ & $0$ & $0$ & $+6$ & $(M-2)/4$ \\ \hline
$4m+3$ & $1$ & $0$ & $0$ & $+3$ & $0$ & $0$ & $+3$ & $(M+1)/4$ \\
&& $1/2$ & $0$ & $+1$ & $+1$ & $+1$ & $+3$ & $(M+1)/4$ \\ \cline{2-9}
& $i$ & $0$ & $+1$ & $0$ & $+1$ & $+1$ & $+3$ & $(M+1)/4$ \\
&& $1/2$ & $+1$ & $+2$ & $0$ & $0$ & $+3$ & $(M+1)/4$ \\ \cline{2-9}
& $-1$ & $0$ & $+2$ & $+1$ & $0$ & $0$ & $+3$ & $(M+1)/4$ \\
&& $1/2$ & $+2$ & $+3$ & $+1$ & $+1$ & $+7$ & $(M-3)/4$ \\ \cline{2-9}
& $-i$ & $0$ & $+3$ & $+2$ & $+1$ & $+1$ & $+7$ & $(M-3)/4$ \\
&& $1/2$ & $+3$ & $0$ & $0$ & $0$ & $+3$ & $(M+1)/4$ \\ \hline
$4m+4$ & $1$ & $0$ & $0$ & $0$ & $0$ & $0$ & $0$ & $M/4+1$ \\
&& $1/2$ & $0$ & $+2$ & $+1$ & $+1$ & $+4$ & $M/4$ \\ \cline{2-9}
& $i$ & $0$ & $+1$ & $+1$ & $+1$ & $+1$ & $+4$ & $M/4$ \\
&& $1/2$ & $+1$ & $+3$ & $0$ & $0$ & $+4$ & $M/4$ \\ \cline{2-9}
& $-1$ & $0$ & $+2$ & $+2$ & $0$ & $0$ & $+4$ & $M/4$ \\
&& $1/2$ & $+2$ & $0$ & $+1$ & $+1$ & $+4$ & $M/4$ \\ \cline{2-9}
& $-i$ & $0$ & $+3$ & $+3$ & $+1$ & $+1$ & $+8$ & $M/4-1$ \\
&& $1/2$ & $+3$ & $+1$ & $0$ & $0$ & $+4$ & $M/4$ \\ \hline
\end{tabular}
}
}
\caption{
The winding number $\chi_{i}$ at the fixed point $p_{i}\ (i=1,2,3,4)$.
All the values of $(M-V_{\eta})/4+1$ exactly agree with the numbers $n_{\eta}$ of the
$\mathbb{Z}_{4}$ physical zero modes given in Table \ref{tab_Z4}.}
\label{Z4tab}
\end{table}

\subsection{$T^2/\mathbb{Z}_6$}
As previously mentioned, there is only a single fixed point under the $\mathbb{Z}_6$ identification $z \sim \omega z$ ($\omega = e^{2 \pi i/6})$, i.e.
\begin{gather}
z = 0 \,\, (\equiv p_1).
\end{gather}
Since the $\mathbb{Z}_6$ group includes its subgroups $\mathbb{Z}_3$ and $\mathbb{Z}_2$, there are additionally two ``$\mathbb{Z}_3$ fixed points" and three ``$\mathbb{Z}_2$ fixed points" that are not invariant under the $\mathbb{Z}_6$ rotation, but invariant under such partial $\mathbb{Z}_3$ and $\mathbb{Z}_2$ rotations up to torus lattice shifts, respectively.
The two $\mathbb{Z}_3$ and three $\mathbb{Z}_2$ fixed points are given by
\begin{align}
&\mathbb{Z}_{3}~\textrm{fixed points:} ~~ 
z = (1 + \tau)/3\,\, (\equiv p_2), \quad 2(1 + \tau)/3 \,\, (\equiv p_3), \\
&\mathbb{Z}_{2}~\textrm{fixed points:} ~~ 
z = 1/2 \,\, (\equiv p_4), \quad \tau/2 \,\, (\equiv p_5), \quad (1 + \tau)/2 \,\, (\equiv p_6).
\end{align}
We should mention that two $\mathbb{Z}_3$ fixed points are exchanged by the $\mathbb{Z}_6$ rotation up to torus lattice shifts, and also that three $\mathbb{Z}_2$ fixed points are connected by the $\mathbb{Z}_6$ rotation.

In a similar way to the previous analyses, we start by considering $\mathbb{Z}_{6}$ eigenstates
\begin{gather}
\xi_\eta (z) = \sum_{\ell=0}^5 \bar \eta^\ell \xi(\omega^\ell z) \qquad (\eta = 1, \hspace{1pt} \omega, \hspace{1pt} \omega^2, \hspace{1pt} \omega^3, \hspace{1pt} \omega^4, \hspace{1pt} \omega^5)\,,
\end{gather}
which belong to the $\mathbb{Z}_{6}$ eigenvalue $\eta = \omega^k$:
\begin{gather}
\xi_{\omega^{k}} (\omega z) = {\omega^{k}} \xi_{\omega^{k}} (z)
 \qquad (k=0,1,\ldots,5).
\end{gather}
We can straightforwardly show the following relations:
\begin{align}
\xi_{\omega^k} (\omega^2 z + \tfrac13 + \tfrac{\tau}3) &= e^{-iq \Lambda_2(z) -2 \pi i (-M/6 + \alpha + 2k/3)} \, \xi_{\omega^k} (z + \tfrac13 + \tfrac{\tau}3), \\
\xi_{\omega^k} (\omega^3 z + \tfrac12) &= e^{-iq \Lambda_1(z) - 2\pi i (\alpha + k/2)} \, \xi_{\omega^k} (z + \tfrac12) \qquad (k=0,1,\ldots,5).
\end{align}
Ignoring the terms related to $\Lambda_1(z)$ and $\Lambda_2(z)$ for infinitesimally small
$|z|$, we obtain
\begin{align}
\xi_{\omega^k} (\omega z) &= \omega^k \xi_{\omega^k} (z), \\
\xi_{\omega^k} (\omega^2 z + \tfrac13 + \tfrac{\tau}3) &= e^{-2 \pi i (-M/6 + \alpha + 2k/3)} \, \xi_{\omega^k} (z + \tfrac13 + \tfrac{\tau}3), \\
\xi_{\omega^k} (\omega^3 z + \tfrac12) &= e^{- 2\pi i (\alpha + k/2)} \, \xi_{\omega^k} (z + \tfrac12) \qquad (k=0,1,\ldots,5).
\end{align}

Suppose that $C_i$ is a sufficiently small contour around the fixed point $p_i$ for each $i$.
Our results of interest are given by using these relations three or two times repeatedly,
\begin{gather}
\chi_1 = k ~~{\rm mod}~6, \label{eq4.63}\\
\chi_2 = \chi_3 = \tfrac{M}2 - 3\alpha - 2k ~~ {\rm mod}~3, \label{eq4.64} \\
\chi_4 = \chi_5 = \chi_6 = -2 \alpha - k ~~ {\rm mod}~2, \label{eq4.65}
\end{gather}
where we have used $\chi_{2}=\chi_{3}$ and $\chi_{4}=\chi_{5}=\chi_{6}$.
Here, the winding number $\chi_{i} \,\, (i=1,2,\ldots,6)$ for $\xi_{\omega^{k}}$
has been defined around the fixed point $p_{i}$ ($i=1,2,\ldots,6$), respectively.

The results in this subsection are summarized in Table \ref{Z6tab}.
We should notice again that although the ``$\mathbb{Z}_3$ and $\mathbb{Z}_2$ fixed
points" are not invariant under the $\mathbb{Z}_{6}$ rotation, zeros at those fixed
points have to be regarded as unremovable ones, and their contribution is
indispensable to guarantee the counting formula \eqref{countingformula}.

We comment on the winding number $\chi_i \,\, (i=1,2,3,4,5,6)$.
As one can see from \eqref{eq4.63}\,--\,\eqref{eq4.65}, $\chi_{1}$ ($\chi_{2, \hspace{.5pt} 3}$ and $\chi_{4, \hspace{.5pt} 5, \hspace{.5pt} 6}$) at the fixed point $p_1$ ($p_{2, \hspace{.5pt} 3}$ and $p_{4, \hspace{.5pt} 5, \hspace{.5pt} 6}$) are less than six (three and two), i.e. $\chi_1 = 0, 1, ..., 5$ ($\chi_{2, \hspace{.5pt} 3} = 0,1,2$ and $\chi_{4, \hspace{.5pt} 5, \hspace{.5pt} 6}= 0,1$).
If an orbifold zero mode wavefunction gives a winding number at $p_1$ ($p_{2, \hspace{.5pt} 3}$ or $p_{4, \hspace{.5pt} 5, \hspace{.5pt} 6}$) larger than or equal to six (three or two), it accidentally contains some contribution from removable zeros.
In other words, some removable zeros accidentally coincide unremovable zeros at the fixed points and then enhance the value of $\chi_i$.
By taking an appropriate linear combination of orbifold zero modes, we can find that the winding number is less than six, three, or two.

\begin{table}[!ht]
\centering
{\tabcolsep = 3.3mm
\renewcommand{\arraystretch}{1.2}
\scalebox{0.85}{
\begin{tabular}{ccc|c:cc:ccc|c|c} \hline
flux & parity & twist & \multicolumn{6}{c|}{winding number} & total & $n_{\eta}$ \\
$M$ & $\eta$ & $\alpha$ & $\chi_1$ & $\chi_2$ & $\chi_3$ & $\chi_4$ & $\chi_5$ & $\chi_6$ & $V_{\eta}=\sum_i \chi_i$ & $(M-V_{\eta})/6+1$ \\ \hline
$6m+1$ & $1$ & $1/2$ & $0$ & $+2$ & $+2$ & $+1$ & $+1$ & $+1$ &$+7$ & $(M-1)/6$\\ 
& $\omega$ & $1/2$ & $+1$ & $0$ & $0$ & $0$ & $0$ & $0$ & $+1$ & $(M+5)/6$\\ 
& $\omega^2$ & $1/2$ & $+2$ & $+1$ & $+1$ & $+1$ & $+1$ & $+1$ & $+7$ & $(M-1)/6$\\ 
& $\omega^3$ & $1/2$ & $+3$ & $+2$ & $+2$ & $0$ & $0$ & $0$ & $+7$ & $(M-1)/6$\\ 
& $\omega^4$ & $1/2$ & $+4$ & $0$ & $0$ & $+1$ & $+1$ & $+1$ & $+7$ & $(M-1)/6$\\ 
& $\omega^5$ & $1/2$ & $+5$ & $+1$ & $+1$ & $0$ & $0$ & $0$ & $+7$ & $(M-1)/6$\\ \hline
$6m+2$ & $1$ & $0$ & $0$ & $+1$ & $+1$ & $0$ & $0$ & $0$ & $+2$ & $(M+4)/6$\\ 
& $\omega$ & $0$ & $+1$ & $+2$ & $+2$ & $+1$ & $+1$ & $+1$ & $+8$ & $(M-2)/6$\\ 
& $\omega^2$ & $0$ & $+2$ & $0$ & $0$ & $0$ & $0$ & $0$ & $+2$ & $(M+4)/6$\\ 
& $\omega^3$ & $0$ & $+3$ & $+1$ & $+1$ & $+1$ & $+1$ & $+1$ & $+8$ & $(M-2)/6$\\ 
& $\omega^4$ & $0$ & $+4$ & $+2$ & $+2$ & $0$ & $0$ & $0$ & $+8$ & $(M-2)/6$\\ 
& $\omega^5$ & $0$ & $+5$ & $0$ & $0$ & $+1$ & $+1$ & $+1$ & $+8$ & $(M-2)/6$\\ \hline
$6m+3$ & $1$ & $1/2$ & $0$ & $0$ & $0$ & $+1$ & $+1$ & $+1$ & $+3$ & $(M+3)/6$\\ 
& $\omega$ & $1/2$ & $+1$ & $+1$ & $+1$ & $0$ & $0$ & $0$ & $+3$ & $(M+3)/6$\\ 
& $\omega^2$ & $1/2$ & $+2$ & $+2$ & $+2$ & $+1$ & $+1$ & $+1$ & $+9$ & $(M-3)/6$\\ 
& $\omega^3$ & $1/2$ & $+3$ & $0$ & $0$ & $0$ & $0$ & $0$ & $+3$ & $(M+3)/6$\\ 
& $\omega^4$ & $1/2$ & $+4$ & $+1$ & $+1$ & $+1$ & $+1$ & $+1$ & $+9$ & $(M-3)/6$\\ 
& $\omega^5$ & $1/2$ & $+5$ & $+2$ & $+2$ & $0$ & $0$ & $0$ & $+9$ & $(M-3)/6$\\ \hline
\end{tabular}
}
}
\caption{
The winding number $\chi_{i}$ at the fixed point $p_{i}\ (i=1,2,3,4,5,6)$.
All the values of $(M-V_{\eta})/6+1$ exactly agree with the numbers $n_{\eta}$ of the
$\mathbb{Z}_{6}$ physical zero modes given in Table \ref{tab_Z6}.}
\label{Z6tab}
\end{table}

\addtocounter{table}{-1}
\begin{table}[!ht]
\centering
{\tabcolsep = 3.3mm
\renewcommand{\arraystretch}{1.2}
\scalebox{0.85}{
\begin{tabular}{ccc|c:cc:ccc|c|c} \hline
flux & parity & twist & \multicolumn{6}{c|}{winding number} & total & $n_{\eta}$ \\
$M$ & $\eta$ & $\alpha$ & $\chi_1$ & $\chi_2$ & $\chi_3$ & $\chi_4$ & $\chi_5$ & $\chi_6$ & $V_{\eta}=\sum_i \chi_i$ & $(M-V_{\eta})/6+1$ \\ \hline
$6m+4$ & $1$ & $0$ & $0$ & $+2$ & $+2$ & $0$ & $0$ & $0$ & $+4$ & $(M+2)/6$\\ 
& $\omega$ & $0$ & $+1$ & $0$ & $0$ & $+1$ & $+1$ & $+1$ & $+4$ & $(M+2)/6$\\ 
& $\omega^2$ & $0$ & $+2$ & $+1$ & $+1$ & $0$ & $0$ & $0$ & $+4$ & $(M+2)/6$\\ 
& $\omega^3$ & $0$ & $+3$ & $+2$ & $+2$ & $+1$ & $+1$ & $+1$ & $+10$ & $(M-4)/6$\\ 
& $\omega^4$ & $0$ & $+4$ & $0$ & $0$ & $0$ & $0$ & $0$ & $+4$ & $(M+2)/6$\\ 
& $\omega^5$ & $0$ & $+5$ & $+1$ & $+1$ & $+1$ & $+1$ & $+1$ & $+10$ & $(M-4)/6$\\ \hline
$6m+5$ & $1$ & $1/2$ & $0$ & $+1$ & $+1$ & $+1$ & $+1$ & $+1$ & $+5$ & $(M+1)/6$\\ 
& $\omega$ & $1/2$ & $+1$ & $+2$ & $+2$ & $0$ & $0$ & $0$ & $+5$ & $(M+1)/6$\\ 
& $\omega^2$ & $1/2$ & $+2$ & $0$ & $0$ & $+1$ & $+1$ & $+1$ & $+5$ & $(M+1)/6$\\ 
& $\omega^3$ & $1/2$ & $+3$ & $+1$ & $+1$ & $+0$ & $+0$ & $+0$ & $+5$ & $(M+1)/6$\\ 
& $\omega^4$ & $1/2$ & $+4$ & $+2$ & $+2$ & $+1$ & $+1$ & $+1$ & $+11$ & $(M-5)/6$\\ 
& $\omega^5$ & $1/2$ & $+5$ & $0$ & $0$ & $0$ & $0$ & $0$ & $+5$ & $(M+1)/6$\\ \hline
$6m+6$ & $1$ & $0$ & $0$ & $0$ & $0$ & $0$ & $0$ &$0$ & $0$ & $M/6+1$\\ 
& $\omega$ & $0$ & $+1$ & $+1$ & $+1$ & $+1$ & $+1$ & $+1$ & $+6$ & $M/6$\\ 
& $\omega^2$ & $0$ & $+2$ & $+2$ & $+2$ & $0$ &$ 0$ & $0$ & $+6$ & $M/6$\\ 
& $\omega^3$ & $0$ & $+3$ & $0$ & $0$ & $+1$ & $+1$ & $+1$ & $+6$ & $M/6$\\ 
& $\omega^4$ & $0$ &  $+4$ & $+1$ & $+1$ & $0$ & $0$ & $0$ & $+6$ & $M/6$\\ 
& $\omega^5$ & $0$ & $+5$ & $+2$ & $+2$ & $+1$ & $+1$ & $+1$ & $+12$ & $M/6-1$\\ \hline
\end{tabular}
}
}
\caption{(Continued.) 
The winding number $\chi_{i}$ at the fixed point $p_{i}\ (i=1,2,3,4,5,6)$.
All the values of $(M-V_{\eta})/6+1$ exactly agree with the numbers $n_{\eta}$ of the
$\mathbb{Z}_{6}$ physical zero modes given in Table \ref{tab_Z6}.}
\end{table}

\subsection{Generic counting formula}
We now turn to a generic zero-mode counting formula on all the orbifolds $T^2/\mathbb{Z}_N$
($N=2,3,4,6$).
Before claiming it, it is convenient to review our ingredients in hand.
The important quantities on the orbifolds $T^2/\mathbb{Z}_N$ are 
given as follows:
\begin{itemize}
\item the flux quanta $M$, where the homogeneous flux $f$ is given as $qf = 2 \pi M$
\item the discretized Scherk-Schwarz twist phase $(\alpha_1, \alpha_2)$
\item the $\mathbb{Z}_N$ eigenvalue $\eta=1,\omega,\ldots,\omega^{N-1}$
($\omega = e^{2\pi i/N}$), where the $\mathbb{Z}_{N}$ eigen zero modes satisfy $\xi_\eta (\omega z) = \eta \, \xi_\eta(z)$
\end{itemize}
These quantities above characterize the orbifold eigen states,
and in fact the numbers of the $\mathbb{Z}_{N}$ eigen zero modes turn out to depend
on $M,\,(\alpha_{1},\alpha_{2})$, $\eta$, and $N$ in a considerably complicated way, as shown
in Tables \ref{tab_Z2}\,--\,\ref{tab_Z6}.

An important quantity here is 
\begin{itemize}
\item the sum of the winding numbers $\chi_{i}$ at the fixed points $p_{i}$ for the $\mathbb{Z}_{N}$ eigenstates belonging to the $\mathbb{Z}_{N}$ eigenvalue $\eta$,
i.e. $V_{\eta} \equiv \sum_i \chi_i$.
\end{itemize}
A complete list of $V_{\eta} = \sum_{i}\chi_{i}$ is ready 
in Table \ref{Z2tab}\,--\,\ref{Z6tab}.
Interesting features that can be read off from the tables are 
\begin{gather}
M - V_{\eta} = 0 ~~ {\rm mod}~N \label{eq4.66}
\end{gather}
and
\begin{gather}
\sum_\eta V_\eta = \sum_{k=0}^{N-1} V_{\omega^{k}} = N^2. \label{eq4.67}
\end{gather}

An important observation is that the quantity
\begin{gather}
\frac{M - V_{\eta}}{N} + 1  \label{eq4.68}
\end{gather}
always takes an integer value even though $M/N$ and 
$V_{\eta}/N$ do not necessarily become integers.
Furthermore, from \eqref{eq4.67}, the quality \eqref{eq4.68} turns out to satisfy
\begin{gather}
\sum_{k=0}^{N-1} \left(\frac{M - V_{\omega^{k}}}{N} + 1 \right) = M. 
 \label{eq4.69}
\end{gather}
Since the number $n_{\omega^{k}}$ of the $\mathbb{Z}_{N}$ eigen zero modes belonging to $\mathbb{Z}_{N}$ eigenvalue $\omega^{k} \,\, (k=0,1,\ldots,N-1)$
satisfies\footnote{
The relation \eqref{eq4.70} comes from the fact that the sum of the 
numbers of all the $\mathbb{Z}_{N}$ eigen zero modes on $T^{2}/\mathbb{Z}_{N}$
is identical to the number of the zero modes on $T^{2}$, i.e. $M$.}
\begin{gather}
\sum_{k=0}^{N-1} n_{\omega^{k}} = M,
 \label{eq4.70}
\end{gather}
the relations \eqref{eq4.69} and \eqref{eq4.70} suggest that the following
equality should hold:
\begin{gather}
n_{\eta} = \frac{M-V_{\eta}}{N}+1.
 \label{eq4.71}
\end{gather}
In fact, we can explicitly verify \eqref{eq4.71} by directly comparing
$n_{\eta}$ in Tables \ref{tab_Z2}\,--\,\ref{tab_Z6} with $(M-V_{\eta})/N+1$
in Tables \ref{Z2tab}\,--\,\ref{Z6tab}.
We call \eqref{eq4.71} a zero-mode counting formula on the magnetized orbifolds $T^2/\mathbb{Z}_N$, and it is the most important
result in this paper.

\section{Discussion and conclusion}
In this paper, we have considered the toroidal orbifolds $T^{2}/\mathbb{Z}_{N}$
($N=2,3,4,6$) with magnetic flux background as 2d extra dimensions.
We have focused on the numbers of the $\mathbb{Z}_{N}$ eigen zero modes on
$T^{2}/\mathbb{Z}_{N}$, which depend on the flux quanta $M$, the SS twist phase $(\alpha_1, \alpha_2)$, and the $\mathbb{Z}_{N}$ eigenvalue $\eta$.
In the previous researches, only a part of such numbers has been obtained,
and neither a generic zero-mode counting formula nor an index theorem
on the orbifolds has been investigated.

In Section 3, we have succeeded in deriving a complete list for the 
numbers of the $\mathbb{Z}_{N}$ eigen zero modes on $T^{2}/\mathbb{Z}_{N}$.
Because of quite complicated dependence on the flux quanta, the SS twist phase, and the $\mathbb{Z}_{N}$ eigenvalue, 
it seems hard that all the numbers of the $\mathbb{Z}_{N}$
eigen zero modes can be universally explained by a simple formula.
Surprisingly, we have found in Section 4 that all the numbers of the $\mathbb{Z}_{N}$
eigen zero modes can be described by a single zero-mode counting formula \eqref{eq4.71}.
A crucial ingredient for the zero-mode counting formula is the sum of the winding numbers at the fixed points on $T^{2}/\mathbb{Z}_{N}$,
i.e. $V_{\eta}$.

Although the origin of the last term in \eqref{eq4.71}
is unclear, the first two terms of $M/N$ and $-V_{\eta}/N$ may be understood from an index theorem point of view, as follows.
From the Atiyah-Singer index theorem, the number of the zero
modes on $T^{2}$ is given by
\begin{gather}
\frac{q}{2\pi}\int_{T^{2}}F = M . \label{eq5.1}
\end{gather}
On the other hand, on the orbifold $T^{2}/\mathbb{Z}_{N}$,
a naive extension of \eqref{eq5.1} would be of the form 
\begin{gather}
\frac{q}{2\pi}\int_{T^{2}/\mathbb{Z}_{N}}F = \frac{M}{N},  \label{eq5.2}
\end{gather}
which may explain the first term $M/N$ in \eqref{eq4.71}.
The reason why $M$ is divided by $N$ in \eqref{eq5.2} is that
the area of the $T^{2}/\mathbb{Z}_{N}$ fundamental domain is given by $(\textrm{the area of}\ T^{2})\times (1/N)$.

An important feature of orbifolds is that they possess fixed points,
which are singularities on manifolds. Hence, they should
be removed from the orbifold fundamental domain.
This observation may explain the second term $-V_{\eta}/N$ in \eqref{eq4.71}.
If the winding number $\chi_{i}$ is non-vanishing at the fixed point $p_{i}$,
it implies the presence of localized flux at the fixed point \cite{Buchmuller:2015eya, Buchmuller:2018lkz}.
That would lead to the second term $-V_{\eta}/N$, because the removal of 
all the fixed points means the subtraction of the localized fluxes at the fixed points from \eqref{eq5.2}.

We have proved the zero-mode counting formula \eqref{eq4.71} by examining the numbers $n_{\eta}$ and $(M-V_{\eta})/N+1$, separately.
It would be of great interest to derive the counting formula \eqref{eq4.71} directly from an index theorem on the orbifolds.
We will pursuit the derivation of our formulae somewhere.

\section*{Acknowledgment}
We would like to thank Shogo Tanimura for important comments at the early stage of the research project.
Y.T. would like to thank Wilfried Buchm\"uller and Markus Dierigl for instructive comments on this manuscript.
M.S. is supported by Japan Society for the Promotion of Science (JSPS) KAKENHI Grant Number JP\,18K03649.
Y.T. is supported in part by Grants-in-Aid for JSPS Overseas Research Fellow (No.\,18J60383) from the Ministry of Education, Culture, Sports, Science and Technology in Japan.

\appendix
\renewcommand{\thesection}{\Alph{section}}

\section{Gamma matrices}
The notation in this paper is basically the same as that in \cite{Abe:2013bca, Abe:2014noa}.
The 6d gamma matrices are taken as
\begin{gather}
\{ \Gamma^M, \Gamma^N \} = 2 \eta^{MN} \qquad (M,N = 0, 1, 2, 3, 5, 6), \\
\eta^{MN} = {\rm diag} \, (+1, -1, -1, -1, -1, -1), \\
\Gamma^\mu = 
\begin{pmatrix}
\gamma^\mu & 0 \\
0 & \gamma^\mu
\end{pmatrix} \qquad (\mu = 0, 1, 2, 3), \\
\Gamma^5 = 
\begin{pmatrix}
0 & i\gamma_5 \\
\gamma_5 & 0
\end{pmatrix}, \qquad 
\Gamma^6 = 
\begin{pmatrix}
0 & \gamma_5 \\
-\gamma_5 & 0
\end{pmatrix}, \qquad 
\Gamma^7 = 
\begin{pmatrix}
\gamma_5 & 0 \\
0 & - \gamma_5
\end{pmatrix}.
\end{gather}
Also, we define
\begin{gather}
\partial_i = \frac{\partial}{\partial y_i} \qquad (i=1,2), \\
\partial = \frac{i}{2 \, {\rm Im} \, \tau} (\bar \tau \partial_1 - \partial_2), \qquad \bar \partial = - \frac{i}{2 \, {\rm Im}\, \tau} (\tau \partial_1 - \partial_2).
\end{gather}

\section{Proof of the generalized Landsberg-Schaar relation}
In this appendix, we give a proof of the generalized Landsberg-Schaar relation
\begin{gather}
\frac1{\sqrt p} \sum_{n = 0}^{p-1} \exp \left( \frac{\pi i (n + \nu)^2 q}{p} \right) = \frac{e^{i\pi/4}}{\sqrt{q}} \sum_{n = 0}^{q-1} \exp\left( -\frac{\pi i n^2 p}{q} - 2 \pi i n \nu \right)
\label{generalizedLS}
\end{gather}
with $p, q \in \mathbb{N}$, $\nu \in \mathbb{Q}$, and $pq + 2 q \nu \in 2\mathbb{Z}$.
Note that \eqref{LS1} and \eqref{LS2} are just special cases of \eqref{generalizedLS}, 
because we can realize them by plugging $\nu = 0$ into the generalized one.

\begin{figure}[!t]
\centering
\includegraphics[width=0.55\textwidth]{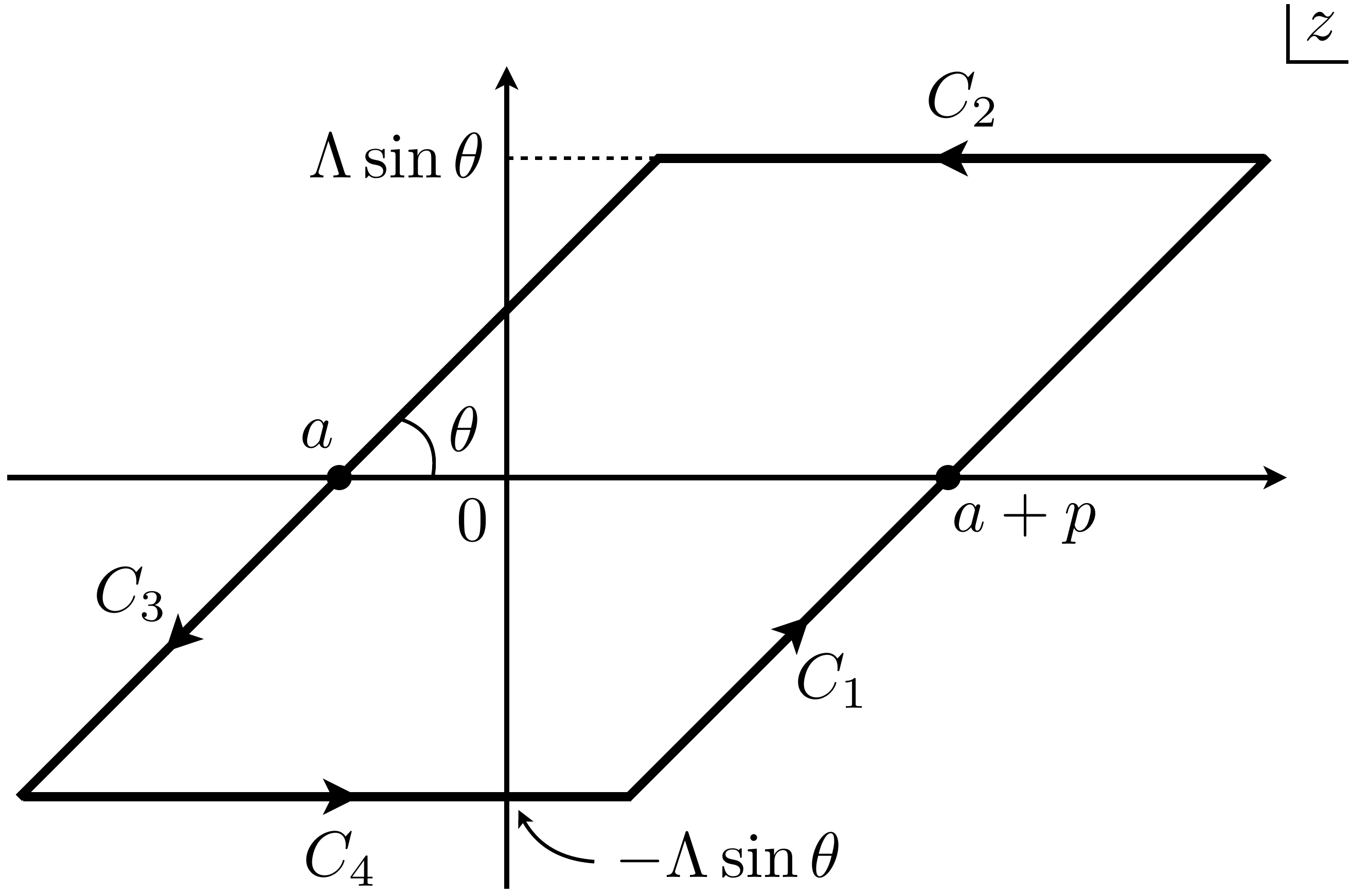}
\caption{A contour that we have adopted.}
\label{LSfig}
\end{figure}

First of all, let us define
\begin{gather}
G(z) = \frac{e^{i \pi q (z + \nu)^2 / p}}{e^{2 \pi i z} - 1}
\end{gather}
and adopt the contour in Figure \ref{LSfig}.
Now, as easily seen, the paths $C_2$ and $C_4$ for $0 < \theta < \pi/4$ do not contribute 
to the integral
\begin{gather}
\int_{C_2, \hspace{1pt} C_4} dz \, G(z) \xrightarrow{\Lambda \to \infty} 0
\end{gather}
in the limit of $\Lambda \to \infty$.

Defining 
\begin{gather}
I = \lim_{\Lambda \to \infty} \bigg\{\int_{C_1} dz \, G(z) + \int_{C_3} dz \, G(z)\bigg\},
\end{gather}
we express the integral $I$ in terms of the new coordinates, $C_1\!:\,z \equiv a + p + re^{i \theta}$ and $C_3\!:\,z \equiv a + re^{i \theta}$ ($-1 < a < 0$), as
\begin{align}
I &= \lim_{\Lambda \to \infty}\int_{-\Lambda}^\Lambda dr \, e^{i\theta} \left[ G(a + p + re^{i\theta}) - G(a + r e^{i\theta}) \right] \notag \\
&= \lim_{\Lambda \to \infty}\int_{-\Lambda}^\Lambda dr \, e^{i\theta} 
\left( \sum_{k=0}^{q-1}e^{2\pi i(a + re^{i\theta})k} \right)
e^{i \pi q (a + re^{i \theta} + \nu)^2 /p}.
\end{align}
Now, by using $x \equiv (re^{i \theta} + a) e^{-i \pi/4}$, we reach
\begin{align}
I &= \lim_{\Lambda \to \infty}
\int_{(-\Lambda e^{i\theta} +a)e^{-i\pi/4}}^{(\Lambda e^{i\theta} +a)e^{-i\pi/4}} 
dx \, e^{i\pi/4} \sum_{k = 0}^{q-1} 
e^{-\pi (q/p) X_k^2 - i\pi (2 \nu k + p k^2 /q )}, \notag \\
&= \int_{-\infty}^{\infty} dx \, e^{i\pi/4} 
\sum_{k = 0}^{q-1} e^{-\pi (q/p) X_k^2 - i\pi (2 \nu k + p k^2 /q )},
\end{align}
where $X_k \equiv x + (\nu + pk/q)e^{-i \pi/4} \,\, (k=0,1, ..., q-1)$.
Performing the Gaussian integrals with respect to $x$ leads to
\begin{align}
I = e^{i \pi/4} \sqrt{\frac{p}{q}} \, \sum_{k=0}^{q-1} e^{-i \pi (2\nu k + p k^2/q)}.
\label{eqB.7}
\end{align}

On the other hand, the residue theorem for the function $G(z)$ gives 
\begin{align}
\oint dz \, G(z) &= 2 \pi i \sum_{k=[a]+1}^{[a] + p} {\rm Res} \, G(z) \notag \\
&= \sum_{k = [a]+1}^{[a] + p} e^{i \pi q (k + \nu)^2 /p},
\end{align}
where $[x] = {\rm max}\,\{n \in \mathbb{Z} \,|\, n \leq x\}$ denotes the floor function.
By imposing $-1 < a < 0$, we finally obtain 
\begin{gather}
\oint dz \, G(z) = \sum_{k=0}^{p-1} e^{i \pi q (k + \nu)^2 /p}.
\label{eqB.9}
\end{gather}
Equating \eqref{eqB.9} with \eqref{eqB.7} yields \eqref{generalizedLS}.
This completes the proof.

\bibliographystyle{unsrt}
\bibliography{references}
\end{document}